\documentclass[12pt]{article}
\usepackage{natbib,graphicx,setspace,lscape,longtable}
\usepackage{natbib,epsfig,graphicx,float}
\usepackage{amsmath,amsthm,amssymb,color}
\usepackage{booktabs} 
\usepackage[nohead]{geometry}
\usepackage[bottom]{footmisc}
\usepackage{comment}
\usepackage{indentfirst}
\usepackage{algorithm}
\usepackage{algorithmic}

\usepackage{endnotes}
\usepackage{verbatim}
\usepackage{threeparttable}
\usepackage{comment}
\usepackage{todonotes}
\usepackage{algorithm}
\usepackage{titlesec}
\usepackage{multibib}
\usepackage{algorithmic}
\usepackage{mathtools}
\usepackage{multirow}
\usepackage{dsfont}
\usepackage{titling}
\usepackage{geometry}
\usepackage{booktabs}
\usepackage{caption}
\usepackage{subcaption}
\usepackage{siunitx}
\usepackage{amsmath}
\usepackage{xcolor}
\usepackage{afterpage}
\usepackage{hyperref}
\RequirePackage[mathlines]{lineno}
\usepackage{multirow}
\usepackage{chngcntr}
\usepackage{caption}
\usepackage{rotating}
\usepackage[titletoc]{appendix}
\bibpunct{(}{)}{;}{a}{,}{,}
\newtheorem{theorem}{Theorem}
\newtheorem{lemma}{Lemma}
\newtheorem{prop}{Proposition}

\floatstyle{ruled}
\newfloat{algorithm}{tbhp}{loa}
\floatname{algorithm}{Algorithm}

\newcommand{\beaa}{\begin{eqnarray*}}
\newcommand{\eeaa}{\end{eqnarray*}}

\newcommand{\datasetshape}[1]{\footnotesize(#1)}

\newcommand{\diag}{\mathrm{diag}}

\newcommand{\argmin}{\mathrm{argmin}}
\newcommand{\vecc}{\mathrm{vec}}

\newcommand{\var}{\mathrm{var}}

\newcommand{\beq}{\begin{eqnarray*}}
\newcommand{\eeq}{\end{eqnarray*}}

\newcites{New}{References}
\titleformat{\section}{\normalfont\Large\bfseries}{\thesection}{0.5em}{}
\titlespacing*{\section} {0pt}{5pt}{3pt}
\titlespacing*{\subsection} {0pt}{5pt}{2pt}
\numberwithin{equation}{section}
\theoremstyle{plain}

\newtheorem{assumption}{Assumption}[section]
\theoremstyle{definition}

\def\ben{\begin{equation*}}
\def\een{\end{equation*}}
\def\bea{\begin{eqnarray}}
\def\eea{\end{eqnarray}}
\def\bean{\begin{eqnarray*}}

\def\eean{\end{eqnarray*}}
\def\bep{\begin{prop}}
\def\eep{\end{prop}}
\def\bc{\begin{center}}
\def\ec{\end{center}}

\def\spacingset#1{\renewcommand{\baselinestretch}%
{#1}\small\normalsize} \spacingset{1}

\renewcommand{\baselinestretch}{1.6}
\allowdisplaybreaks[4]
\title{Tucker Diffusion Model for High-dimensional Tensor Generation\thanks{All authors have contributed equally.}}
\author{Guo, J., Kong, X., Li, Z. \& Mao, J.}

\begin{document}

\maketitle
\begin{abstract}
Statistical inference on large-dimensional tensor data has been extensively studied in the literature and widely used in economics, biology, machine learning, and other fields,  but how to generate a structured tensor with a target distribution is still a new problem. As profound AI generators, diffusion models have achieved remarkable success in learning complex distributions. However, their extension to generating multi-linear tensor-valued observations remains underexplored. In this work, we propose a novel Tucker diffusion model for learning high-dimensional tensor distributions. We show that the score function admits a structured decomposition under the low Tucker rank assumption, allowing it to be both accurately approximated and efficiently estimated using a carefully tailored tensor-shaped architecture named \texttt{Tucker-Unet}. Furthermore, the distribution of generated tensors, induced by the estimated score function, converges to the true data distribution at a rate depending on the maximum of tensor mode dimensions, thereby offering a clear theoretical advantage over the naive vectorized approach, which has a product dependence. Empirically, compared to existing approaches, the Tucker diffusion model demonstrates strong practical potential in synthetic and real-world tensor generation tasks, achieving comparable and sometimes even superior statistical performance with significantly reduced training and sampling costs.

\end{abstract}

\noindent%
\bigskip

\spacingset{1.7} 

\section{Introduction}

With the rapid advancement of data acquisition technology, multi-array (tensor) data has found extensive applications in various fields, including economics, biology, and machine learning. Statistical inference for such data, often formulated as tensor sequences, has become fundamental to identifying and modeling correlation structures both within and across different modes. This is evidenced by a series of seminal works, such as \cite{wang2019factor,chen2021statistical,Yu2021Projected,he2024online,gao2023two, chen2022factor, chang2023modelling,han2024tensor,auddy2025tensors,chen2024semi}.
However, a significant open question remains: how can we estimate the underlying distribution of structured tensors? Going a step further, how can we generate synthetic tensors that accurately follow a target distribution? A promising approach is to leverage the Tucker decomposition as a foundational structure for the data and to employ deep generative models as a primary tool for sampling. The integration of these two concepts—originating from tensor algebra and deep learning—presents substantial technical challenges. Consequently, the theoretical insights arising from this synthesis are anticipated to be profound.

The intersection of structured tensor modeling (via Tucker decomposition) and generative modeling (like deep generative models) represents a significant frontier for statistical inference on multi-array data. While existing literature excels at correlation analysis and modeling within tensor sequences, estimating the underlying distribution of structured tensors and subsequently generating synthetic counterparts remains a critical gap.
A Tucker-decomposition-based generative framework offers a principled approach. The core tensor and loading matrices define a structured, lower-dimensional manifold. A deep generative model (e.g., diffusion model) can be trained to learn the joint distribution of these decomposed components from real tensor data. 
The technical challenges are profound: designing architectures that respect multi-linear algebraic constraints, ensuring identifiability in the generative process, and scaling to high-order tensors. The potential theoretical insights—such as characterizing the learnability of tensor-valued distributions, the sample complexity for generative tasks, and the statistical guaranties of synthetic data—would substantially extend the current inference paradigm, enabling reliable data augmentation, hypothesis testing, and simulation for complex multi-way data structures.

In a mathematical sense, deep generative models (DGMs) are deep neural networks that learn high-dimensional probability distributions, e.g, those of images, texts, and audio, to generate new samples that resemble the initial dataset \citep{lai2025principlesdiffusionmodels}. The learned distributions can then be used for further statistical inference, privacy protection, and numerical simulation purposes, e.g., creating a high-fidelity environment in which reinforcement learning agents interact and improve performance \citep{chen2024overview, han2024lifelike}.

The diffusion model, one of the state of the art DGMs, achieves tremendous success in various fields of interest, including computer vision, computational biology, and reinforcement learning \citep{sohl2015deep,song2019generative,ho2020denoising,song2020score,chen2024overview}. In the forward process of a diffusion model, an observation is sequentially corrupted by injected random noise and eventually turns into white noise. In the backward process, a trained denoising network is able to gradually remove the noise and generate samples of high perceptual quality from scratch. The denoising network is closely related to the score function, i.e., the gradient of the log-probability density \citep{vincent2011connection}. 

Compared to other DGMs, such as generative adversarial networks (GANs) \citep{creswell2018generative,goodfellow2020generative} and variational auto-encoders (VAEs) \citep{kingma2013auto,kingma2019introduction}, the multi-step score-based models typically achieve better performance in generation quality \citep{song2020score,chen2024overview}. On the other hand, the higher generation fidelity of diffusion models comes at the cost of heavy computational burdens in the backward sampling process, as the score function, often approximated by a large neural network\footnote{In our experiments, a typical diffusion model that learns matrices of shape $64\times 64$ has more than 1 billion parameters.}, needs to be evaluated dozens, and sometimes even hundreds of times \citep{song2020denoising,karras2022elucidating}.

While most existing theories for diffusion models focus on vector-valued initial data distributions \citep{chen2022sampling,chen2023score,chen2025diffusion}, in data-intensive domains such as healthcare and finance, it is increasingly common to observe data sources that are naturally structured as tensors, which are multi-way arrays capturing complex dependencies across different modes. Examples include, but are not limited to, multi-modal remote sensing, medical images, macroeconomic panel data \citep{wang2019factor,chen2021statistical, Yu2021Projected,athey2021matrix,choi2024matrix}, virtual drug screening for predicting gene expression profiles \citep{radhakrishnan2022simple}, multiple networks and graphs \citep{zheng2022limit,morris2020tudataset,helma2001predictive}, and multiple-input multiple-output (MIMO) radars \citep{sun2015mimo}.

The key challenge is that many tensors are naturally high-dimensional. As a quick reminder, for an order-$3$ tensor with dimension $p_d=100$ in each mode, simple vectorization will result in a vector of dimension $p = 1,000,000$, which poses a heavy burden for numerical computation \citep{wang2019factor,chen2021statistical, choi2024matrix} . Meanwhile, the number of observed tensors is often limited in many applications. For instance, the number of networks (e.g., yearly international trading networks all over the world) in a multiple network dataset is typically comparable to, and sometimes even much smaller than, the network dimensions, leading to the classical small $n$ large $p$ problem in high-dimensional statistics \citep{negahban2012unified,wainwright2019high}. Hence, it is crucial, both theoretically and practically, to \emph{tailor low-dimensional structures for diffusion models that learn high-dimensional tensor distributions}, for the sake of data-efficient training and accelerated sampling under sample size and computational constraints.

\subsection{Initial tensor data distribution}

To achieve dimension reduction for diffusion models, the pioneering papers \cite{chen2023score,chen2025diffusion} work under the assumption that the data points are approximately supported on an unknown low-dimensional linear subspace. The authors claim that, with a properly chosen neural network structure, the score function can be accurately approximated by a neural network and efficiently estimated through the training process. In this work, we shall instead assume that observed high-dimensional tensors lie approximately within the \emph{Tucker manifold}, defined as
\begin{equation*}
\mathcal{M}_{\text{Tucker}} := \left\{ 
X \in \mathbb{R}^{p_1 \times p_2 \times \cdots \times p_D} 
\;\middle|\; 
X = F \times_{d=1}^D A_d, \;
A_d \in \mathrm{St}(r_d, p_d), \;
F \in \mathbb{R}^{r_1 \times r_2 \times \cdots \times r_D}
\right\},
\end{equation*}
where $\mathrm{St}(r_d, p_d) = \{ A_d \in \mathbb{R}^{p_d \times r_d} \mid A_d^\top A_d = I_{r_d} \}$ is the Stiefel manifold of orthonormal matrices. To this end, this work can also be viewed as a budget-friendly method for generating tensors from this specific manifold, saving both training and sampling costs. In particular, we assume the observed data tensors $\{\mathbf{X}^{i}_0\}_{i=1}^{n}$ approximately have the following low-rank Tucker structure. For $i\in [n]:=\{1,\dots,n\}$,
\begin{equation}\label{tensor factor model}
\prod_{d=1}^{D}p_{d}^{\beta_{d}/2} \times \mathbf{X}^i_{0} = \mathbf{F}^i\times_{d=1}^{D}\Lambda_{d}+\mathbf{E}^i,
\end{equation}
where $\mathbf{F}^i\in \mathbb{R}^{r_{1}\times\ldots\times r_{D}}$ denotes the core tensor; for $A_d \in \mathrm{St}(r_d, p_d)$, $\Lambda_{d}=p_{d}^{\beta_{d}/2}A_{d}\in \mathbb{R}^{p_{d}\times r_{d}}$ represents the factor loading matrix associated with the $d$-th mode; $\mathbf{E}^i\in \mathbb R^{p_{1}\times\ldots\times p_{D}}$ is the tensor of idiosyncratic residuals; and $D$ and $\{r_{d}\}_{d\in [D]}$ are fixed positive integers. Here, the scaling of $\mathbf{X}^i_{0}$, together with Assumption \ref{assum:2}, is meant to ensure a constant signal strength so that the initial distribution projected onto the core space is neither degenerated nor divergent, and the resulting forward diffusion within the core is regular. The subscript $0$ in $\mathbf{X}^i_{0}$ denotes the starting time of the forward diffusion process. We say that the tensor $\mathbf{X}^{i}_0$ follows the initial tensor data distribution $P_0$. The primary statistical goal is to learn $P_0$ in a statistically efficient way and to generate new tensors from the learned distribution with low computational costs.

\subsection{Contributions}

\begin{figure}[h]
  \centering
    \includegraphics[width=0.75\linewidth]{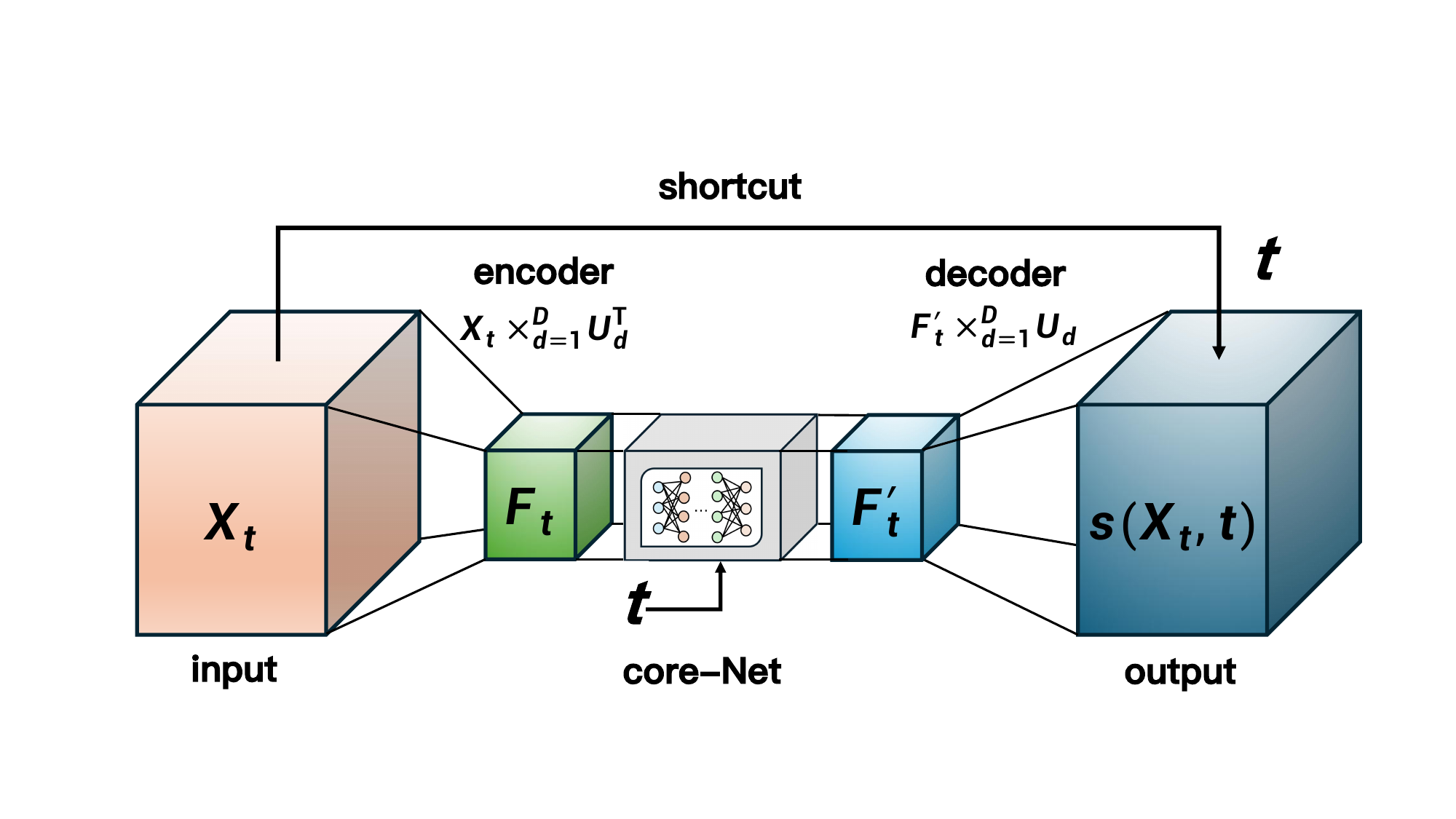}
  \caption{Visualization of the homogeneous \texttt{Tucker-Unet} by deactivating the noise heterogeneity structure. The details of \texttt{Tucker-Unet} can be found in Section \ref{sec:tuckerunet}. The proposed network consists of a Tucker encoder and decoder, a core-Net and a shortcut skip connection towards the output.}
  \label{fig:tuckerunet}
\end{figure}

The main contributions of this work are as follows. First, we introduce the Tucker diffusion model, a principled framework for learning high-dimensional tensor-valued distributions. Central to our approach is a novel structural decomposition of the score function under the low Tucker rank assumption, which enables the development of a specialized tensor-shaped deep architecture, \texttt{Tucker-Unet}, integrating the Tucker encoder and decoder, a low-dimensional score network, and a shortcut skip connection. Rooted in the growing literature on multi-way structural modeling in high-dimensional statistics \citep{wang2019factor,Yu2021Projected,chen2021statistical,yuan2023two}, the \texttt{Tucker-Unet} explicitly exploits the inherent multi-linear structure of tensors, mitigating the dimensionality challenges inherent in vectorization-based methods. A schematic illustration (Figure \ref{fig:tuckerunet}) roughly elucidates the information flow and architectural components of \texttt{Tucker-Unet}, see Section \ref{sec:tuckerunet} for details. 

By utilizing the multi-linear structure, our model provides a scalable solution for high-dimensional tensor generation. To some extent, the proposed Tucker diffusion model can be viewed as the tensor analog of the latent space diffusion methods designed for vector distributions \citep{vahdat2021score,rombach2022high,jing2022subspace}. Instead of first projecting data onto some pre-specified latent spaces and then learning the low-dimensional diffusion model, \texttt{Tucker-Unet} has a built-in low-dimensional Tucker structure that ensures efficient high-dimensional end-to-end training and sampling, avoiding the necessity of multi-stage or cascading generation \citep{ho2022cascaded,jing2022subspace}.

While many existing diffusion models also deal with tensor-structured data, such as images and videos, to the best of our knowledge, these models do not take into account the intrinsic multi-way linear low-rank structures. Instead, they rely more on local feature extractors, such as convolutional neural networks (CNNs), which were initially designed for image classification purposes \citep{ho2020denoising,song2020score,chen2024overview}. As a conceptual counter-example of local feature extractors in our cases, note that, unlike images, for macroeconomic panels, global trade networks, and protein graphs, the columns and rows per channel can switch arbitrarily without losing any initial information. As also shown by the numerical experiments, it is more natural to harness the multi-way low-rank structures using \texttt{Tucker-Unet} for a number of applications discussed later in this work \citep{wang2019factor,radhakrishnan2022simple,zheng2022limit}.

Then, we also establish theoretical guaranties for the proposed Tucker diffusion model. By developing approximation and estimation theories for the score function, we prove that the convergence rate of the generated tensor distribution to the true data distribution depends on the maximum of tensor mode dimensions, instead of the product of tensor mode dimensions via the naive vectorized approach. That is, our theoretical results highlight the fundamental efficiency of exploiting tensor structure, as the proposed \texttt{Tucker-Unet} is able to provably mitigate the curse of dimensionality in the case of high-order tensors.

Finally, we validate the practical efficacy of \texttt{Tucker-Unet} through comprehensive experiments on generating synthetic and real-world tensors. To the best of our knowledge, this is the first empirical attempt to realize scalable diffusion-based tensor generation by explicitly harnessing the multi-linear low-rank structures, built upon prior theoretical treatments of low-dimensional linear models \citep{chen2023score,chen2025diffusion}. Empirical results demonstrate that \texttt{Tucker-Unet} matches or surpasses state-of-the-art methods in statistical performance metrics while achieving up to a 98\% reduction in sampling time. These findings establish the method’s feasibility for applications demanding both generative fidelity and computational efficiency in high-dimensional tensor generation.

\subsection{Organization and notations}

The remainder of this article is organized as follows. In Section \ref{sec:method}, we introduce the Tucker diffusion model for high-dimensional tensor generation, which is based on a novel score decomposition lemma and the resulting \texttt{Tucker-Unet} design. In Section \ref{sec:theory}, we provide the theoretical rates concerning score approximation, score estimation, and distribution convergence, thereby demonstrating the advantages of utilizing multi-linear low-rank structures. Then, in Section \ref{sec:syn} and Section \ref{sec:real}, we conduct extensive numerical experiments on generating synthetic and real-world tensors, where the proposed Tucker diffusion model consistently outperforms its competitors in terms of comparable generative fidelity under much smaller training and sampling costs. Finally, in Section \ref{sec:discussion}, we discuss possible future directions.

To conclude this section, we introduce some notations used throughout the paper. For constants $a$ and $b$,  $a = \mathcal{O}(b) $ indicates that $a$ is bounded above by $b$ up to a constant factor, while $a = \Theta(b)$ means $b = \mathcal{O}(a)$; $a\sim b$ implies that $a $ and $b$ are of the same order, i.e., $a = \mathcal{O}(b)$ and $a = \Theta(b)$. For a vector $\mathbf{a}$, we let $\|\mathbf{a}\|_2$ and $\|\mathbf{a} \|_{\infty}$ be its $\ell^2$ and $\ell^{\infty}$ norms, respectively. For a matrix $\mathbf A$, we denote $\mathrm{tr}(\mathbf A)$, $\|\mathbf A\|_{F}$, and $\|\mathbf A\|$ as its trace, Frobenius norm, and operator norm, respectively. When $\mathbf{A}$ is symmetric, we denote $\lambda_{\max}(\mathbf{A)}$, $\lambda_k(\mathbf{A})$, and $\lambda_{\min}(\mathbf{A)}$ as the largest, the $k$-th largest, and the smallest eigenvalues of $\mathbf{A}$, respectively. For a tensor $\mathbf A$, we let $\|\mathbf A\|_{F}$ be its Frobenius norm. For two tensors $\mathbf{A}$ and $\mathbf{B}$, we define $\mathbf{A}\oslash \mathbf{B}$ as the element-wise division of $\mathbf{A}$ by $\mathbf{B}$. For a random tensor $\mathbf{X}$ following the distribution $P$, we write $\|\mathbf X\|_{L^2(P)}^2 = \mathbb{E}\left(\|\mathbf X\|_{F}^2\right)$. Finally, we denote $\phi(\cdot; \mu, \Sigma)$ as the Gaussian density function with mean $\mu$ and covariance $\Sigma$. 

\section{Tucker Diffusion Model}\label{sec:method}
In this section, we introduce the Tucker diffusion model that learns high-dimensional tensor distributions. A generative diffusion model typically consists of a forward process and a backward process. The forward process progressively adds noise to the original data. In the same spirit as the vector-valued diffusion models \citep{song2020score,chen2023score,chen2024overview}, we consider the tensor-valued Ornstein-Uhlenbeck process, which is described by the following SDE,
\begin{equation}\label{T1}
\mathrm{d} \mathbf{X}_t=-\frac{1}{2} v_{t} \mathbf{X}_t\mathrm{d} t+\sqrt{v_{t}} \mathrm{d} \mathbf{W}_t\quad\text{with}~\mathbf{X}_0 \sim P_{0}~\text{and}~t\in[0, T],
\end{equation}
where $\{\mathbf{W}_t\}_{t \geq 0}$ is a standard tensor-valued Wiener process, $v_{t}>0$ is a non-decreasing weighting function, $T$ is a terminal time, and $P_{0}$ is the initial data distribution. We denote $P_t$ as the marginal distribution of $\mathbf{X}_t$ with a corresponding density function $p_t$. Given an initial tensor $\mathbf{X}_0$, at time $t$, the conditional distribution of $\mathbf{X}_t | \mathbf{X}_0$ is Gaussian, i.e.,
\begin{equation}\label{T2}
\mathbf{X}_t | \mathbf{X}_0\sim \mathcal{N}\left(\alpha_t \mathbf{X}_{0}, \left\{(1-\alpha_t^2)^{1/D} I_{p_{d}}\right\}_{d=1}^{D}\right),
\end{equation}
where $\alpha_t=\exp (-\int_0^t v_{s}\mathrm{d} s/2)\in(0,1]$ is the shrinkage ratio, and $1-\alpha_t^2$ is the variance scale of the added Gaussian noise. For simplicity, we shall take $v_t=1$ in this work. Note that the terminal distribution $P_T$ is close to $P_{\infty}=\mathcal{N}(0, \{I_{p_{d}}\}_{d=1}^{D})$ when $T$ is sufficiently large, since the marginal distribution of an Ornstein-Uhlenbeck process converges exponentially fast to its stationary distribution.

\begin{figure}[h]
  \centering
    \includegraphics[width=0.75\linewidth]{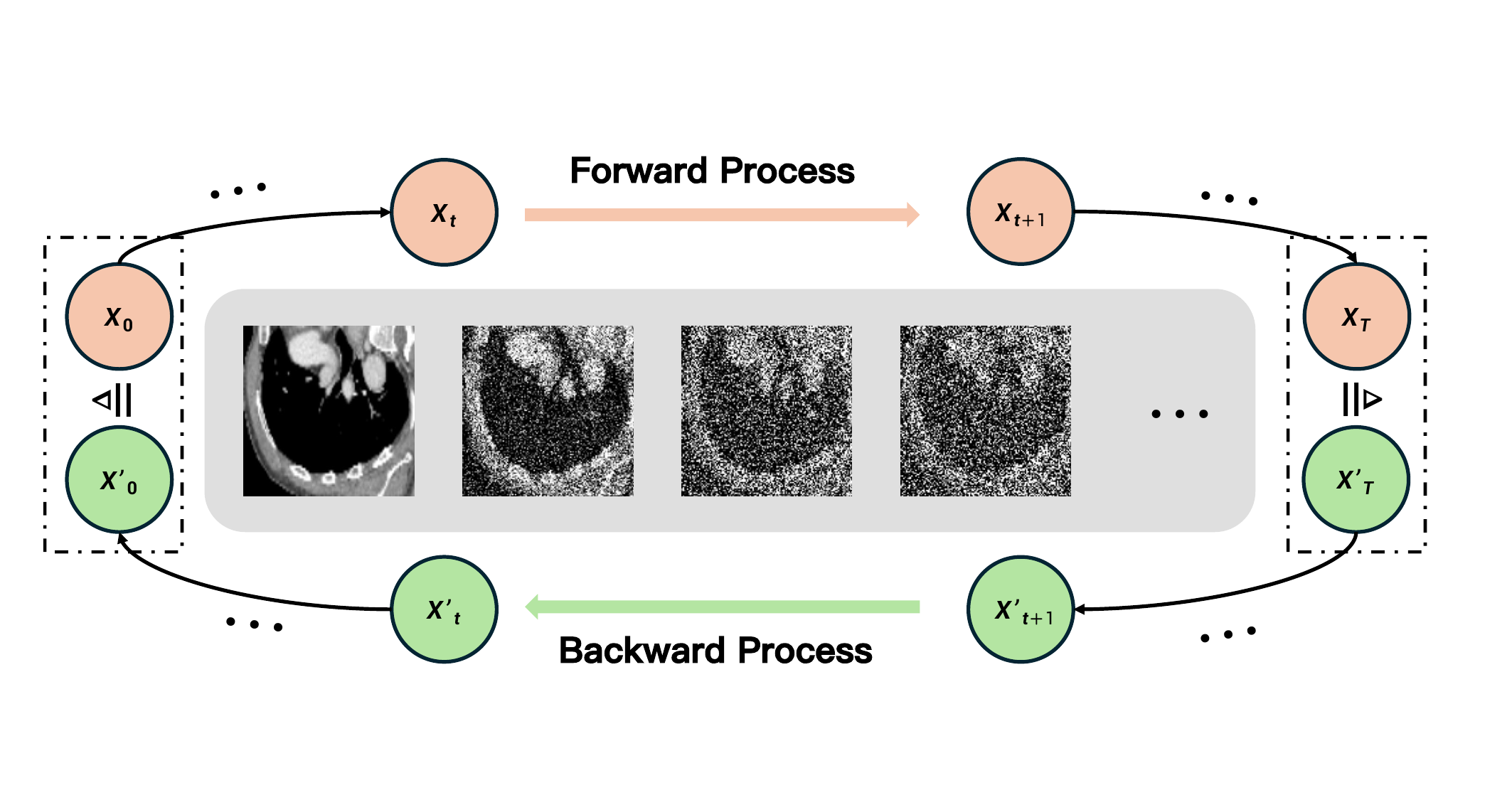}
  \caption{A visual demonstration of the forward and backward processes of a diffusion model using a real-world medical image.}
  \label{fig:demo}
\end{figure}

To generate new samples from the initial data distribution $P_0$, we can reverse the forward process in time, which yields a well-defined backward process that transforms noise into data. We write the time-reversed SDE associated with (\ref{T1}) as
\begin{equation}\label{T3}
\mathrm{d} \mathbf{X}_t^{\leftarrow}=\left(\frac{1}{2} \mathbf{X}_t^{\leftarrow}+\nabla \log p_{T-t}\left(\mathbf{X}_t^{\leftarrow}\right)\right) \mathrm{d} t+\mathrm{d} \overline{\mathbf{W}}_t \quad \text {with}~\mathbf{X}_0^{\leftarrow} \sim Q_0 \text { and } t \in[0, T]
\end{equation}
where $\{\overline{\mathbf{W}}_t\}_{t \geq 0}$ is another Wiener process independent of $\{\mathbf{W}_t\}_{t \geq 0}$, and $Q_0$ is the initial distribution of the backward process. Note that $\nabla \log p_t(\cdot)$, the tensor-valued score function, is unknown in real world applications. Hence, a deep neural network is learned to approximate the score function, which we denote as 
$$\hat{\mathbf{S}}\left(\widehat{\mathbf{X}}_t^{\leftarrow}, T-t\right)\approx \nabla \log p_{T-t}\left(\mathbf{X}_t^{\leftarrow}\right).$$
With $\hat{\mathbf{S}}$ and $Q_{0}=\mathcal{N}(0, \{I_{p_{d}}\}_{d=1}^{D})$ that is easy to sample\footnote{For tensor-valued Ornstein-Uhlenbeck processes, the error introduced by taking $Q_0=\mathcal{N}(0, \{I_{p_{d}}\}_{d=1}^{D})\approx P_T$ decays exponentially with respect to $T$.}, we have the following implementable process for data generation
\begin{equation}\label{T4}
\mathrm{d} \widehat{\mathbf{X}}_t^{\leftarrow}=\left(\frac{1}{2} \widehat{\mathbf{X}}_t^{\leftarrow}+\hat{\mathbf{S}}\left(\widehat{\mathbf{X}}_t^{\leftarrow}, T-t\right)\right) \mathrm{d} t+\mathrm{d} \overline{\mathbf{W}}_t\quad\text{with}~\widehat{\mathbf{X}}_0^{\leftarrow} \sim\mathcal{N}\left(0, \{I_{p_{d}}\}_{d=1}^{D}\right).
\end{equation}

\subsection{Score matching}

To estimate the score function, a conceptual approach is to minimize a weighted quadratic loss; that is,
\begin{equation}\label{T5}
\hat{\mathbf{S}} = \argmin_{\mathbf{S} \in \mathcal{S}} \int_0^T \mathbb{E}_{\mathbf{X}_t \sim P_t}\left[\left\|\nabla_{\mathbf{X}_{t}} \log p_t\left(\mathbf{X}_t\right)-\mathbf{S}\left(\mathbf{X}_t, t\right)\right\|_{F}^2\right] \mathrm{d} t
\end{equation}
where $\mathcal{S}$ is a concept class, often a set of deep neural networks to be specified later. However, such an objective function is intractable, as $\nabla\log p_t$ cannot be observed in practice. We can instead minimize an equivalent objective \citep{song2020score,chen2024overview}, i.e.,
\begin{equation}\label{T6}
\hat{\mathbf{S}} = \argmin_{\mathbf{S} \in \mathcal{S}} \int_0^T \mathbb{E}_{\mathbf{X}_0 \sim P_{ 0}}\left[\mathbb { E }_{ \mathbf { X }_{t} | \mathbf { X }_{0}} \left[\left\| \nabla_{\mathbf{X}_t} \log \Phi_t\left(\mathbf{X}_t | \mathbf{X}_0\right)-\mathbf{S}\left(\mathbf{X}_t, t\right) \right\|_{F}^2\right]\right] \mathrm{d} t
\end{equation}
Here $\Phi_t(\mathbf{X}_t | \mathbf{X}_0)$ is the transition kernel of the forward tensor OU process, which has an analytical form:
\begin{equation}\label{T7}
\nabla_{\mathbf{X}_t} \log \Phi_t\left(\mathbf{X}_t | \mathbf{X}_0\right)=-\frac{\mathbf{X}_t-\alpha_t \mathbf{X}_0}{1-\alpha_t^2}.
\end{equation}
Indeed, $\nabla_{\mathbf{X}_t} \log \Phi_t\left(\mathbf{X}_t | \mathbf{X}_0\right)$ is the noise added to $\mathbf{X}_0$ at time $t$. Therefore, \eqref{T6} is also known as denoising score matching, as $\hat{\mathbf{S}}(\mathbf{X}_t, t)$ approximates the injected noise at time $t$. 

During the training process, given $\mathbf{X}_{0}^{1},\ldots,\mathbf{X}_{0}^{n} \overset{\text{i.i.d.}}{\sim} P_{0}
$, we sample $\mathbf{X}_t$ given $\mathbf{X}_0=\mathbf{X}_{0}^{i}$ from $\mathcal{N}(\alpha_t \mathbf{X}^{i}_{0}, \{(1-\alpha_t^2)^{1/D} I_{p_{d}}\}_{d=1}^{D})$, where $t$ is sampled uniformly from the interval $\left[t_0, T\right]$\footnote{Here, $t_0$ is an early-stopping time to prevent the score function from blowing up as $t\rightarrow 0$ \citep{song2019generative,chen2023score}.} for some small $t_0>0$. As such, the empirical score matching objective is
\begin{equation}\label{T8}
\tilde{\mathbf{S}} = \argmin_{\mathbf{S} \in \mathcal{S}} \widehat{\mathcal{L}}(\mathbf{S})= \argmin_{\mathbf{S} \in \mathcal{S}}  \frac{1}{n} \sum_{i=1}^n \ell\left(\mathbf{X}_{0}^{i} ; \mathbf{S}\right),\quad\text{for}
\end{equation}
\begin{equation*}
    \ell\left(\mathbf{X}_{0}^{i} ; \mathbf{S}\right):= \frac{1}{T-t_0} \int_{t_0}^T \mathbb{E}_{\mathbf{X}_t | \mathbf{X}_0=\mathbf{X}_{0}^{i}}\left[\left\|\nabla_{\mathbf{X}_t} \log \Phi_t\left(\mathbf{X}_t | \mathbf{X}_0\right)-\mathbf{S}\left(\mathbf{X}_t, t\right)\right\|_{F}^2\right] \mathrm{d} t. 
\end{equation*} 
Note that we have implicitly assumed sufficient sampling of $\mathbf{X}_t | \mathbf{X}_0$ and $t$, as they are easy to generate given $ \mathbf{X}_0$. As a result, the randomness in $\widehat{\mathcal{L}}(\mathbf{S})$ arises solely from finite initial tensor samples from $P_0$. We shall denote the population loss as $\mathcal{L}(\mathbf{S}):=\mathbb{E}_{P_{0}}[\widehat{\mathcal{L}}(\mathbf{S})]$.

\subsection{Score decomposition and \texttt{Tucker-Unet}}\label{sec:tuckerunet}

If the initial tensor observations $\{\mathbf{X}^{i}_0\}_{i=1}^{n}$ have the low-rank Tucker structure as given in \eqref{tensor factor model}, we expect the score function to exhibit a specific multi-linear structure that can lead to a more efficient neural network architecture. To be more specific, we now introduce a score function decomposition from the tensor time-reversed SDE in \eqref{T3}. From this score decomposition, a \texttt{Tucker-Unet} structure is designed to model and learn the score function, which enables more data-efficient training and accelerated sampling. Throughout this work, we focus on the following low-rank tensor model for the initial tensor distribution $P_0$.

\begin{assumption}[Low-rank Tucker factor model]\label{assum:1}
We assume the following conditions on the low-rank tensor model \eqref{tensor factor model}.
    \begin{itemize}
        \item[(i)] For each mode $d\in [D]$, the factor loading matrix satisfies $\Lambda_{d}^{\top}\Lambda_{d}/p_{d}^{\beta_{d}}=I_{r_{d}}$, i.e., its columns are orthonormal after proper scaling.
        \item[(ii)] Let $\mathbf{f}=\vecc(\mathbf{F})\in \mathbb{R}^{r}$, the factor $\mathbf{f}$ is assumed to follow a distribution with density $p_{\mathrm{core}}$ that has zero mean and a finite second moment.
        \item[(iii)]  The random noise $\mathbf{E}$ is independent of $\mathbf{F}$ and follows a tensor normal distribution $\mathcal{N}\left(0,\{\Sigma_{e}^{d}\}_{d\in [D]}\right)$ with $\Sigma_{e}^{d}=\diag(\sigma^{2}_{d1},\ldots,\sigma^{2}_{dp_{d}})$. In addition, there exist positive constants $\sigma_{\min}^{2}$ and $\sigma_{\max}^{2} $ such that $$0\leq\sigma_{\min}^{2}\leq\min_{d\in[D]}\min_{j\in [p_{d}]}\sigma_{dj}^{2}\leq\max_{d\in[D]}\max_{j\in [p_{d}]}\sigma_{dj}^{2}\leq\sigma_{\max}^{2}<\infty.$$   
    \end{itemize}
\end{assumption}

Assumption~\ref{assum:1}~(i) is standard in the literature on tensor factor models; see for example Assumption 2 of \cite{barigozzi2025statisticalinferencelargedimensionaltensor}. In particular, our signal condition generally applies to both strong and weak factors, depending on the value of $\beta_d\in[0,1]$, where larger $\beta_d$ means larger signal level on the $d$-th mode. Assumption~\ref{assum:1}~(ii) imposes finite second moment of core tensor factor, which is also standard for factor models; see Assumption A in \cite{bai2003inferential}. Finally, Assumption~\ref{assum:1}(iii) is similar to Assumption 1 in \cite{chen2025diffusion} to ensure a  well-defined score decomposition. Note that we allow heterogeneity in the idiosyncratic noise, which brings additional technical challenges.

Now, define $p^{\beta}=\prod_{d=1}^{D} p^{\beta_{d}}$ with $p=\prod_{d=1}^{D}p_{d}$ and $\beta\in [0,1]$,  $r=\prod_{d=1}^{D}r_{d}$, $\Sigma_{e}=\otimes_{d=D}^{1}\Sigma_{e}^{d}=\diag(\sigma_{1}^{2},\ldots,\sigma_{p}^{2})\in \mathbb{R}^{p\times p}$, $\Sigma_{t}=h_{t}I_{p}+\alpha_{t}^{2}p^{-\beta}\Sigma_{e}\in \mathbb{R}^{p\times p}$, 
$$\Sigma_{t}^{\mathrm{Tucker}}=\mathrm{reshape}\left\{\diag(\Sigma_{t});\{p_{d}\}_{d\in[D]}\right\}\in \mathbb{R}^{p_{1}\times\ldots\times p_{D}},$$ $\Sigma_{A_t}=(A^{\top}\Sigma_{t}^{-1}A)^{-1}\in \mathbb{R}^{r\times r}$ for $A=\otimes_{d=D}^{1} A_{d}\in \mathbb{R}^{p\times r}$, and $$\mathbf{g}_{t}=\Sigma_{A_{t}}\vecc\left((\mathbf{X}_{t}\oslash\Sigma_{t}^{\mathrm{Tucker}})\times_{d=1}^{D}A_{d}^{\top}\right)\in \mathbb{R}^{r\times 1},$$the ground-truth score function has the following decomposition.

\begin{lemma}[Score decomposition lemma]\label{lem: Heterogeneity}
Under Assumption \ref{assum:1}, the score function $\nabla\log p_t(\mathbf{X}_{t})$ can be decomposed into a subspace score and a complement score as
\begin{align}\label{eq:iso_decomp}
\nabla \log p_t(\mathbf{X}_{t})=&\underbrace{\left(\mathrm{Tucker}\left(\Sigma_{A_{t}}\nabla\log p_{\textrm{core}}^{t}\left(\mathbf{g}_{t}\right)\right)\times_{d=1}^{D}A_{d}\right)\oslash \Sigma_{t}^{\mathrm{Tucker}}}_{\mathrm{subspace~score}}-\nonumber\\
&\underbrace{\left(\mathbf{X}_{t}-\mathrm{Tucker}\left(\mathbf{g}_{t}\right)\times_{d=1}^{D}A_{d}\right)\oslash\Sigma_{t}^{\mathrm{Tucker}}}_{\mathrm{complement~score}}.
\end{align}
where $p_{\mathrm{core}}^{t}(\mathbf{g}_{t})\coloneqq\int\phi\left(\mathbf{g}_{t}; \alpha_t\mathbf{f},\Sigma_{A_{t}}\right) p_{\mathrm{core}}(\mathbf{f}) \mathrm{d}\mathbf{f}$ and $\mathrm{Tucker}(\cdot)\coloneqq\mathrm{reshape}\left\{\cdot;\{r_{d}\}_{d\in [D]}\right\}$.
\end{lemma}

To provide a clearer picture of the score function within the core space, we now analyze a Gaussian example. Let $\Sigma_{e}=\sigma^{2}I_{p}$ and the vectorized factor follow the Gaussian distribution $\mathbf{f}\sim\mathcal{N}\left(0,\Sigma_{f}\right)$ with $\Sigma_{f}=\diag(\sigma_{f_1}^{2},\ldots,\sigma_{f_r}^{2})$ and $r=\prod_{d=1}^{D}r_{d}$. Defining $\Sigma_{\mathbf{g}_{t}}^{\mathrm{Tucker}}=\mathrm{Tucker}\left(\mathbf{g}_{t}\right)$ for $\Sigma_{\mathbf{g}_{t}}=\diag\left(h_{t}+\sigma^{2}\alpha_{t}^{2}p^{-\beta}+\alpha_{t}^{2}\sigma_{f_1}^{2},\ldots,h_{t}+\sigma^{2}\alpha_{t}^{2}p^{-\beta}+\alpha_{t}^{2}\sigma_{f_r}^{2}\right)$, we have
 \begin{align}\label{Gauusian score}
   \nabla \log p_t(\mathbf{X}_{t})=&\underbrace{-\left[\left(\mathbf{X}_{t}\times_{d=1}^{D}A_{d}^{\top}\right)\oslash\Sigma_{\mathbf{g}_{t}}^{\mathrm{Tucker}}\right]\times_{d=1}^{D}A_{d}}_{\mathrm{subspace~score}}-\nonumber\\
   &\underbrace{\frac{1}{h_{t}+\sigma^{2}\alpha^{2}_{t}p^{-\beta}}(\mathbf{X}_{t}-\mathbf{X}_{t}\times_{d=1}^{D}A_{d}A_{d}^{\top})}_{\mathrm{complement~score}},
 \end{align}

Essentially, Lemma \ref{lem: Heterogeneity} suggests massive dimensionality reduction for the score function under the low-rank Tucker structure. To see this, we can rearrange the terms in \eqref{eq:iso_decomp} and obtain
\begin{equation}\label{eq:tucker-unet}
 \begin{aligned}
\mathbf{g}_{t}&=\Sigma_{A_{t}}\vecc\left[\left(\underbrace{\mathbf{X}_{t}\oslash\Sigma_{t}^{\mathrm{Tucker}}}_{1. \text{FiLM}}\right)\underbrace{\times_{d=1}^{D}A_{d}^{\top}}_{2. \text{Tucker Encoder}}\right],\\
\nabla\log p_t(\mathbf{X}_t)&=\bigg[\underbrace{\mathrm{Tucker}\left(\xi\left(\mathbf{g}_{t},t\right)\right)}_{3. \texttt{Core-Net}}\underbrace{\times_{d=1}^{D}A_{d}}_{4. \text{Tucker Decoder}}\underbrace{-\mathbf{X}_t\bigg]\oslash\Sigma_{t}^{\mathrm{Tucker}}}_{5. \text{Skip Connection}},
 \end{aligned}
\end{equation}
where $\xi\left(\mathbf{g}_{t}, t\right)=\Sigma_{A_{t}}\nabla\log p_{\mathrm{core}}^{t}\left(\mathbf{g}_{t}\right)+\mathbf{g}_{t}$ is a low-dimensional function that characterizes the core diffusion process, which is typically approximated by a deep neural network. 

Most interestingly, the five components in (\ref{eq:tucker-unet}) build a concrete neural network architecture. Now we explain the layers from inside to outside. 
In the first part of the forward propagation, \texttt{Tucker-Unet} utilizes an optional \footnote{The convolutional FiLM layer is a standard technique used to account for the noise heterogeneity structure characterized by $\Sigma_{t}^{\mathrm{Tucker}}$. In fact, this module is theoretically unnecessary under homogeneous noise or strong signal dominance. In some of our numerical experiments involving strong signals, deactivating this block improves training stability.} convolutional Feature-wise Linear Modulation (FiLM) layer \citep{perez2018film} and a novel multi-way Tucker encoder to map the input $\mathbf{X}_t$ to $\mathbf{g}_t$. In the second part, the low-dimensional core diffusion is approximated by the \texttt{Core-Net}. Third, we remap the low-dimensional core back to the high-dimensional tensor space using a multi-way Tucker decoder. Finally, the shortcut skip connection with temporal correction can be approximated by the final convolutional residual block.

\begin{algorithm}
\caption{Pseudo-code for \texttt{Tucker-UNet} forward pass}
\label{alg:tuckerunet}
\begin{algorithmic}[1]
\REQUIRE Tensor $\mathbf{X} \in \mathbb{R}^{p_1 \times \cdots \times p_D}$, time step $t$;

\STATE $\mathbf{Z} \gets \text{FiLMConv}(\mathbf{X},t)$ \hfill \COMMENT{Feature-wise Linear Modulation (optional)}

\STATE $\mathbf{G} \gets \mathbf{Z} \times_{d=1}^{D}U^{\top}_{d}$ \hfill \COMMENT{Tucker Encoding}

\STATE $\mathbf{G}' \gets \texttt{Core-Net}(\mathbf{G}; t)$ \hfill \COMMENT{Score in the core space}

\STATE $\mathbf{Z}' \gets \mathbf{G}' \times_{d=1}^{D}U_{d}$ \hfill \COMMENT{Tucker Decoding}

\STATE $\mathbf{Y} \gets \text{Conv} \circ  \text{ResBlock}(\mathbf{Z}', \mathbf{Z}, t)$ \hfill \COMMENT{Shortcut skip connection from Step 1}

\RETURN Score prediction $\mathbf{Y}\in \mathbb{R}^{p_1 \times \cdots \times p_D}$.
\end{algorithmic}
\end{algorithm}

In Algorithm \ref{alg:tuckerunet}, we present the pseudo-code of the proposed \texttt{Tucker-UNet}. The \texttt{Core-Net} in the core space can be parameterized using a deep neural network with ReLU activations, which takes low-dimensional inputs and is computationally efficient. Meanwhile, the other components of \texttt{Tucker-UNet}, e.g., the convolutional FiLM layer, the Tucker encoder and decoder, and the shortcut skip connection, are mostly elementary algebraic operations and can be well-approximated using light-weight neural networks. 

To obtain the theory, we regularize the class of score functions $\mathcal{S}_{\mathrm{Tucker}}$ as
\begin{eqnarray*}
&& \mathcal{S}_{\mathrm{Tucker}}\nonumber\\
&=&\left\{\mathbf{S}_{U, \Omega, \boldsymbol{\theta}}(\mathbf{X}_{t}, t)=\left(\mathbf{\zeta}_{\boldsymbol{\theta}}^{\mathrm{Tucker}}\left(\Sigma_{U_{t}}\vecc\left(\left(\mathbf{X}_{t}\oslash\Omega_{t}^{\mathrm{Tucker}}\right)\times_{d=1}^{D}U_{d}^{\top}\right),t\right)\times_{d=1}^{D}U_{d}-\mathbf{X}_{t}\right)\oslash\Omega_{t}^{\mathrm{Tucker}}\right\},
\end{eqnarray*}
where the $U$ parameters $\{U_d \in \mathbb{R}^{p_d \times r_d}; d=1,..., D\}$ satisfy $U_{d}^{\top}U_{d}=I_{r_{d}}$ and $\Sigma_{U_{t}}=\left(U^{\top}\Omega_{t}^{-1}U\right)^{-1}$ with $U=\otimes_{d=D}^{1} U_{d}$;
the low-dimensional ReLU network parameterized by $\theta$ is a function
$$
\mathbf{\zeta}^{\mathrm{Tucker}}_{\boldsymbol{\theta}}: \mathbb{R}^{r_{1}\times\cdots\times r_{D}}\times\left[t_{0}, T\right] \rightarrow \mathbb{R}^{r_{1}\times\cdots\times r_{D}};
$$
and the $\Omega$ parameters 
$
\{\Omega_{t}^{\mathrm{Tucker}}=\mathrm{reshape}\left\{\diag(\Omega_{t});\{p_{d}\}_{d\in [D]}\right\}; t_0\leq t\leq T\}
$
where $\Omega_{t}=\left(\otimes_{d=D}^{1}\Omega^{d}\right)\times \alpha_{t}^{2}p^{-\beta}+h_{t}I_{p}$ with $\Omega^{d}=\diag(\omega_{d1},\ldots, \omega_{dp_{d}})\in [0, \sigma_{\max}^{2}]^{p_{d}}$. 

We configure the low-dimensional core ReLU network $\vecc\left(\mathbf{\zeta}^{\mathrm{Tucker}}_{\boldsymbol{\theta}}\right) =\mathbf{\zeta}_{\boldsymbol{\theta}}\in\mathcal{F}_{\mathrm{core}}$ using hyper-parameters, where $\mathcal{F}_\mathrm{core}\left(L, M, J, K, \kappa, \gamma_{\mathbf{g}}, \gamma_t\right)$ is a family of neural networks defined as
$$\begin{aligned} 
& \mathcal{F}_\mathrm{core}\left(L, M, J, K, \kappa, \gamma_{\mathbf{g}}, \gamma_t\right)\\ =&\Bigg\{\mathbf{\zeta}_{\boldsymbol{\theta}}(\mathbf{g}_{t}, t)=W_L \sigma\left(\ldots \sigma\left(W_1\left(\mathbf{g}_{t}^{\top}, t\right)^{\top}+\mathbf{b}_1\right) \ldots\right)+\mathbf{b}_L~\text{with}~\boldsymbol{\theta}=\left\{\mathbf{W}_{l},\mathbf{b}_{l}\right\}_{l\in [L]}: \\ 
&\text{network width bounded by}~M,\quad\sup _{\mathbf{g}_{t}, t}\|\mathbf{\zeta}_{\boldsymbol{\theta}}(\mathbf{g}_{t}, t)\|_{2} \leq K, \\ 
&\max \left\{\left\|\mathbf{b}_l\right\|_{\infty},\left\|W_l\right\|_{\infty}\right\} \leq \kappa~\text {for}~l\in [L], \quad\sum_{l=1}^L\left(\left\|W_l\right\|_0+\left\|\mathbf{b}_l\right\|_0\right) \leq J, \\ 
&\left\|\mathbf{\zeta}_{\boldsymbol{\theta}}\left(\mathbf{g}_{t1}, t\right)-\mathbf{\zeta}_{\boldsymbol{\theta}}\left(\mathbf{g}_{t2}, t\right)\right\|_{2} \leq \gamma_{\mathbf{g}}\left\|\mathbf{g}_{t1}-\mathbf{g}_{t2}\right\|_{2}~\text{for any}~t \in[t_{0}, T], \\ 
&\left\|\mathbf{\zeta}_{\boldsymbol{\theta}}\left(\mathbf{g}_{t}, t_1\right)-\mathbf{\zeta}_{\boldsymbol{\theta}}\left(\mathbf{g}_{t}, t_2\right)\right\|_{2} \leq \gamma_t\left|t_1-t_2\right|~\text{for any}~\mathbf{g}_{t}\Bigg\}.\end{aligned}$$
where the ReLU activation function is applied entrywise, computing $ \mathrm{ReLU}(a) = \max\{a, 0\} $. The $ \{W_l\}_{l \in [L]} $ and $ \{b_l\}_{l \in [L]} $ represent the weight matrices and intercepts, respectively. Correspondingly, the width and depth of the network are denoted by $ M = \max_{l \in [L]} \dim(\mathbf{b}_l) $ and $ L $, respectively. Here, the uniform bound $ \sup_{\mathbf{g}_t, t} \| \mathbf{\zeta}_{\boldsymbol{\theta}}(\mathbf{g}_t, t) \|_2 \leq K $ and the sparsity constraint $ \sum_{l=1}^L \left( \|W_l\|_0 + \|\mathbf{b}_l\|_0 \right) \leq J $ are standard assumptions for ReLU networks \citep{bartlett2017spectrally, song2020score,chen2023score, chen2025diffusion}. The Lipschitz continuity of $ \zeta_{\theta} $ is often enforced by Lipschitz network training \citep{gouk2021regularisation} or induced by the implicit bias of the training algorithm \citep{bartlett2020benign}.

\subsection{Backward sampling}
Finally, after learning the score function using the score matching technique given in \eqref{T8}, with either the default high-dimensional network structures or the proposed low-dimensional \texttt{Tucker-UNet}, we can utilize the DDIM sampler from \cite{song2020denoising}, which is a notably fast and stable diffusion sampling scheduler \citep{ma2025efficient}, to generate new tensors.

When generating samples with a diffusion model, the total time consumption can be calculated via (NFE, number of function evaluations) $\times$ (time per function evaluation). This work primarily focuses on reducing the (time per function evaluation) by tailoring dimension reduction for the score function when generating high-dimensional tensors of low Tucker ranks. Indeed, reducing the NFE given any learned score function is also an active field of research, i.e., the training-free sampling approaches \citep{karras2022elucidating,ma2025efficient}. However, this topic is far beyond the main purpose of this work, and we shall leave it for future pursuits.

\section{Theory}\label{sec:theory}
In this section, given the score decomposition and the class of the score network $\mathcal{S}_{\mathrm{Tucker}}$, we establish three theoretical guaranties. Section \ref{subsec: score approximation} establishes an approximation theory for $\mathcal{S}_{\mathrm{Tucker}}$, showing that it can approximate any score function of the form \eqref{eq:iso_decomp} under suitable hyper-parameter choices. Section \ref{subsec: score estimation} establishes generalization bounds of finite-samples for the estimation of the score. Section \ref{subsec: distribution estimation} establishes distributional convergence by bounding the total variation distance between the tensor distributions.

\subsection{Theory of Score Approximation}\label{subsec: score approximation}
In this subsection, we develop an approximation theory for the score function. For notational convenience, we decompose the score function into an on-support subspace component and an off-support complement component. Specifically, we define the subspace score as
$$\mathbf{S}_{\mathrm{sub}}(\mathbf{g}_{t},t)=\left(\mathrm{Tucker}\left(\Sigma_{A_{t}}\nabla\log p_{\mathrm{core}}^{t}\left(\mathbf{g}_{t}\right)\right)\times_{d=1}^{D}A_{d}\right)\oslash \Sigma_{t}^{\mathrm{Tucker}},$$
and the complement score as 
$$\mathbf{S}_{\mathrm{com}}(\mathbf{X}_{t},t)=\left(\mathbf{X}_{t}-\mathrm{Tucker}\left(\mathbf{g}_{t}\right)\times_{d=1}^{D}A_{d}\right)\oslash\Sigma_{t}^{\mathrm{Tucker}}.$$
Here, $\mathbf{S}_{\mathrm{sub}}$ captures the score component restricted to the low-rank Tucker subspace, while $\mathbf{S}_{\mathrm{com}}$ accounts for deviations orthogonal to this subspace. To establish the score approximation accuracy, we impose the following assumptions on the core factor distribution $P_{\mathrm{core}}$.
\begin{assumption}[Core tensor factor distribution] \label{assum:2}
    The density function $p_{\mathrm{core}}$ is non-negative and twice continuously differentiable. We define $\Sigma_{f}=\var(\mathbf{f})$ and $0<\lambda_{\min}\left(\Sigma_{f}\right)\leq\lambda_{\max}\left(\Sigma_{f}\right)<\infty$.
    Furthermore, assume there exist positive constants $B, C_1$ and $C_2$ such that
   \begin{equation}\label{eq:subgaussian}
    p_{\mathrm{core}}(\mathbf f) \leq (2\pi)^{-\frac{r}{2}}C_1 \exp(-C_2\|\mathbf f\|_2^2 /2)\quad \text{when}\quad\|\mathbf f\|_2 \ge B.
\end{equation}
\end{assumption}

Assumption \ref{assum:2} characterizes the tail behavior of $P_{\mathrm{core}}$ as sub-Gaussian, which is commonly adopted in the high-dimensional statistics literature \citep{vershynin2018high, wainwright2019high} and also in recent works on diffusion model theory \citep{chen2023score, chen2025diffusion}. We also impose the following regularity assumption on the score function.

\begin{assumption}[Lipschitz continuity of low-rank tensors]\label{assum:3}
The on-support score function $\mathbf{S}_{\mathrm{sub}}(\mathbf{g}, t)$ is $L_{\mathbf{g}}$-Lipschitz in $\mathbf{g} \in \mathbb{R}^{r}$ for any $t \in[t_{0}, T]$.
\end{assumption}

The Lipschitz assumption on the score function is standard in the diffusion model literature \citep{chen2022sampling, chen2025diffusion, han2024neural}. As a concrete example, the Gaussian case in \eqref{Gauusian score} with $L_{\mathbf{g}}= \max\left\{\sigma_{fr}^{-2},1\right\}$ satisfies Assumption \ref{assum:3}. Under these conditions, we can characterize the approximation error between the true score function and its low-rank surrogate, as formalized in the following theorem.
\begin{theorem}[Tucker score approximation]\label{theorem:1}
Given an approximation error $\epsilon>0$, we can choose the configuration of the network architecture $\mathcal{F}_{\mathrm{core}}$ such that
\begin{gather*}
M=\mathcal{O}\left((1+L_{\mathbf{g}})^{r}\log^{r/2}(r/(t_{0}\epsilon))TL_{t}\epsilon^{-(r+1)}\right),
\quad \gamma_{\mathbf{g}}=10r(L_{\mathbf{g}}+1)(1+\sigma_{\max}^{2D}), \\
K=\mathcal{O}\left((1+L_{\mathbf{g}})\log^{1/2}\left(r/(t_{0}\epsilon)\right)\right),\quad L=\mathcal{O}\left(\log(K/\epsilon)+1\right),\quad
\gamma_{t}=10L_{t},\\
J=\mathcal{O}\left((1+L_{\mathbf{g}})^{r}\log^{r/2}(r/(t_{0}\epsilon))TL_{t}\epsilon^{-(r+1)}\left(\log( K/\epsilon)+1\right)\right),\\
\kappa=\max\{(1+L_{\mathbf{g}})(1+\sigma_{\max}^{2D})\log^{1/2}(r/(t_{0}\epsilon)),TL_{t}\},
\end{gather*}
where 
\begin{align*}
L_{t}&=\sup_{\|\mathbf{g}\|_{\infty}\leq\sqrt{\log(r/(t_{0}\epsilon))}}\sup_{t\in [t_{0},T]}\left\|\frac{\partial}{\partial t} \mathbf{\xi}(\mathbf{g}, t)\right\|_2=\mathcal{O}\left(\frac{(L_{\mathbf{g}}+1)\log^{3/2}(r/(t_{0}\epsilon))}{t_{0}^{2}}\right).
\end{align*}
Then for any data distribution $P_{0}$ satisfying Assumptions \ref{assum:1}, \ref{assum:2}, \ref{assum:3}, there exists a network $\mathbf{S}_{\bar{U},\bar{\Omega},\bar{\boldsymbol{\theta}}} \in \mathcal{S}_{\mathrm{Tucker}}$ such that for any $t \in[t_{0}, T]$, we have
\begin{align*}
\left\|\mathbf{S}_{\bar{U},\bar{\Omega},\bar{\boldsymbol{\theta}}}(\mathbf{X},t)-\nabla \log p_t(\mathbf{X})\right\|_{L^2\left(P_t\right)}\leq \frac{\sqrt{r} + 1}{h_{t}} \epsilon.
\end{align*}
\end{theorem}

The proof of Theorem \ref{theorem:1} is deferred to Appendix B. Some discussions are stated as follows. First, we provide an $L^2$-approximation error bound over the unbounded input domain $\mathbb{R}^{p_{1}\times\ldots\times p_{D}}$, addressing the unbounded-ness problem through a careful truncation argument, where time $t$ is incorporated as an additional input. Second, the approximation error exhibits benign dependence on the dimension, primarily depending on $r=\prod_{d=1}^{D}r_d$ rather than $p=\prod_{d=1}^{D}p_{d}$: the network size is determined solely by the core subspace dimension. The reason for this is that the Tucker encoder-decoder module has $0$ approximation error under our model assumptions, as ensured by the key score decomposition in Lemma \ref{lem: Heterogeneity}.

\subsection{Theory of Score Estimation}\label{subsec: score estimation}
In this subsection, we establish the sample complexity for score estimation using $\mathcal{S}_{\mathrm{Tucker}}$. Since we have parameterized the score function using deep neural networks, we can express the score matching objective in \eqref{T8} as 
\begin{equation*}\label{eq:empirical loss}
\mathbf{S}_{\widehat{U},\widehat{\Omega},\widehat{\theta}}=\arg\min_{\mathbf{S}_{U,\Omega,\boldsymbol{\theta}}\in \mathcal{S}_{\mathrm{Tucker}}}  \frac{1}{n(T-t_{0})} \sum_{i=1}^n  \int_{t_0}^T \mathbb{E}_{\mathbf{X}_t | \mathbf{X}_0=\mathbf{X}_{0}^{i}}\left(\left\|\mathbf{S}_{U,\Omega,\boldsymbol{\theta}}\left(\mathbf{X}_t, t\right)+\frac{\mathbf{X}_t-\alpha_{t}\mathbf{X}_0^{i}}{h_{t}}\right\|_{F}^2\right)\mathrm{d}\mathbf{t}.
\end{equation*}
Given $\mathbf{X}_{0}^{1},\ldots,\mathbf{X}_{0}^{n} \overset{\text{i.i.d.}}{\sim} P_{0}
$ and $p_{\max}=\max\{p_{1},\ldots, p_{D}\}$, the following theorem establishes the $L^2$ convergence of $\mathbf{S}_{\widehat{U},\widehat{\Omega},\widehat{\theta}}(\cdot,t)$ to $\nabla \log p_t(\cdot)$ when the sample size $n$ grows to infinity.
\begin{theorem}[Tucker score estimation]\label{theorem:2}
Suppose Assumptions \ref{assum:1}, \ref{assum:2}, \ref{assum:3} hold, if we take $\tau_{n}=\frac{(r+12)\log \log n}{2\log n}$ and $\epsilon=n^{-\frac{1-\tau_{n}}{r+5}}$ in Theorem \ref{theorem:1}, then with probability $1 - 1/n$,
    \begin{align*}
      &\frac{1}{p^{1-\beta}(T-t_0)}\int_{t_0}^T \mathbb{E}_{P_{t}}\left(\left\|\mathbf{S}_{\widehat{U},\widehat{\Omega},\widehat{\theta}}(\mathbf{X}_{t}, t) - \nabla \log p_t(\mathbf{X}_{t})\right\|_{F}^{2}\right)\mathbf{d}\mathbf{t}\nonumber\\
      =&\mathcal{O}\left(\left(\frac{1}{t_{0}^{5}}+\frac{T}{t_{0}^{2}}\right)\left(n^{-\frac{2-2\tau_{n}}{r+5}}+p_{\max}n^{-\frac{r+3+2\tau_{n}}{r+5}}\right)(1+L_{\mathbf{g}})^{r+6}\log^{r/2+6}(1/t_{0})\log T\log^{2}p_{\max}\log^{3}n\right).
    \end{align*}
\end{theorem}
In Theorem \ref{theorem:2}, we provide an explicit sample complexity bound for score matching, with the proof deferred to Appendix C. However, as $ t_0 \to 0 $, the theorem becomes vacuous due to the blowup of the score function $ \nabla \log p_t $. While a larger $ t_0 $ improves the estimation error bound, extending the backward process to a large $ t_0 $ leads to poor distribution recovery. Mathematically, the distribution error $\mathrm{TV}(P_{t_0}, \widehat{P}_{t_0})$ increases as $t_0$ decreases, due to the increasing score estimation error. On the other hand, the core generation bias, measured by $\mathrm{W}_2(P_{\mathrm{core}}^{t_0}, P_{\mathrm{core}})$, is of order $\mathcal{O}(\sqrt{t_0})$ and therefore decreases to zero as $t_0 \to 0$, where $p_{\mathrm{core}}^{t}$ is the density of the distribution $P_{\mathrm{core}}^{t}$.. This reveals a tradeoff in the recovery of the latent factor tensor. Although we cannot directly translate the total variation distance to the Wasserstein-2 distance, or vise versa, to assess the recovery of the core tensor factors, \cite{chen2023score, hu2024statistical} suggests that the choice of $t_0 \sim n^{-\frac{1-\tau_{n}}{3(r+5)}}$ strikes a balance between the early stopping error and the score estimation error, enabling the consistent recovery, up to some orthogonal transformations, of the latent core tensor.

\subsection{Theory of Distribution Estimation}\label{subsec: distribution estimation}

This section provides the distribution estimation guaranties based on the estimated score function. Recall that, in practice, diffusion models generate data through the backward process. Given the estimated score function $\mathbf{S}_{\widehat{U},\widehat{\Omega},\widehat{\theta}}$ as stated in Theorem \ref{theorem:2}, we denote the generated distribution using $\mathbf{S}_{\widehat{U},\widehat{\Omega},\widehat{\theta}}$ as $\widehat{P}_{t_0}$. The following theorem demonstrates that the simulated distribution is accurate with high probability.
\begin{theorem}[Tensor distribution estimation]\label{theorem:3}
Suppose Assumptions \ref{assum:1}, \ref{assum:2}, and \ref{assum:3} hold. Given the neural score estimator $\mathbf{S}_{\widehat{U},\widehat{\Omega},\widehat{\theta}}\in \mathcal{S}_{\mathrm{Tucker}}$ in Theorem \ref{theorem:2}, we choose $t_{0}=\mathcal{O}\left(n^{-\frac{1-\tau_{n}}{3(r+5)}}\right)$ and $T\sim\log n+\log p_{\max}$. Then, with probability $1 - 1/n$, it holds that
\begin{align*}
p^{-\frac{1-\beta}{2}}\mathrm{TV}(P_{t_{0}}, \widehat{P}_{t_0})=&\mathcal{O}\left((1+L_{\mathbf{g}})^{r/2+3}\left(n^{-\frac{1-\tau_{n}}{6(r+5)}}+\sqrt{p_{\max}}n^{-\frac{3r+4+11\tau_{n}}{6(r+5)}}\right)\log^{2}p_{\max}\log^{r/4+6}n\right).
\end{align*}
\end{theorem}
The proof is given in Appendix D. Theorem \ref{theorem:3} has the following interpretations. First, the $p^{(1-\beta)/2}$ factor on the left-hand side of Theorem \ref{theorem:3} originates from our scaling of data that ensures a constant core signal strength and, hence, a regular forward diffusion process within the core space. The rescaled sample complexity bound is, in fact, independent of the dimension $p=\prod_{d=1}^{D}p_{d}$, which is consistent with Theorem 3 of \cite{chen2023score}.

\begin{table}[h]
\caption{A comparison of our theoretical result to those from the closely related works in terms of tensor distribution estimation, under the model assumptions of this work. The results for Chen et al. (2023) and (2025b) are dervied by vectorizing the tensors since their theories work only for vector sequences.}
\label{tab:distribution estimation literature}
\centering
\resizebox{1\textwidth}{!}{%
\begin{tabular}{lcccc}
\toprule
$\mathrm{TV}(P_{t_{0}}, \widehat{P}_{t_0})$ & Signal regime & $\beta=0$  & $\beta=1$\\
\midrule
\cite{chen2023score} &$\beta = 1$    &  not established  &   $n^{-\frac{1-\tau_{n}}{6(r+5)}}+\sqrt{p}n^{-\frac{3r+4+11\tau_{n}}{6(r+5)}}$\\
\cite{chen2025diffusion} &$\beta = 0$ &  $p^{1/2}\left(n^{-\frac{1-\tau_{n}}{6(r+5)}}+\sqrt{p}n^{-\frac{3r+4+11\tau_{n}}{6(r+5)}}\right)$    & not established    \\
\texttt{Tucker-Unet}&$\beta\in[0,1]$     & $p^{1/2}\left(n^{-\frac{1-\tau_{n}}{6(r+5)}}+\sqrt{p_{\max}}n^{-\frac{3r+4+11\tau_{n}}{6(r+5)}}\right)$  & $n^{-\frac{1-\tau_{n}}{6(r+5)}}+\sqrt{p_{\max}}n^{-\frac{3r+4+11\tau_{n}}{6(r+5)}}$\\
\bottomrule
\end{tabular}
}
\label{tab:comparison}
\end{table}

Second, recall that the Tucker subspace is a special low-dimensional linear subspace with a Kronecker structure \citep{wang2019factor,Yu2021Projected,chen2021statistical}. If we disregard the Tucker structure and vectorize the tensors, based on previous results on low-dimensional diffusion models \citep{chen2023score,chen2025diffusion}, we can derive the corresponding results in Table \ref{tab:distribution estimation literature}. In Theorem \ref{theorem:3}, if we keep the Tucker structure in mind, the sample complexity bound can be improved from $\sqrt{p}n^{-\frac{3r+4+11\tau_{n}}{6(r+5)}}$ to $\sqrt{p_{\max}}n^{-\frac{3r+4+11\tau_{n}}{6(r+5)}}$, for $p=\prod_{d=1}^{D}p_{d}$. That is, given a finite number of modes, the convergence rate depends on the maximum of the tensor mode dimensions, in contrast to the product dependence induced by naive vectorization. This markedly mitigates the curse of dimensionality associated with high-order tensor data.

\section{Numerical Simulation}\label{sec:syn}
In this section, we conduct numerical simulation experiments to demonstrate the practical benifit of the proposed Tucker diffusion model using \texttt{Tucker-Unet}. For synthetic data, we generate high-dimensional observations based on a matrix factor model, which is commonly used to model panels from finance and macroeconomics and captures structured bilinear variation through a small number of latent factors; note that matrices are order-2 tensors \citep{wang2019factor,chen2021statistical,choi2024matrix}.
     
The observed data are matrix-valued observations $\mathbf{X}_0^i \in \mathbb{R}^{p_1 \times p_2}$ generated from a low-rank latent structure driven by a smaller latent factor matrix $\mathbf{F}^i \in \mathbb{R}^{r_1 \times r_2}$, with loading matrices governing the row and column dependencies. Specifically, each observation $\mathbf{X}_0^i$ is generated according to the following model:
\begin{equation*}
    \mathbf{X}_0^i = R \mathbf{F}^i C^\top + \mathbf{E}^i, \quad i \in [n],
\end{equation*}
where $R \in \mathbb{R}^{p_1 \times r_1}$ and $C \in \mathbb{R}^{p_2 \times r_2}$ are the row and column factor loading matrices, respectively; $\mathbf{F}^i$ is the latent factor matrix; and $\mathbf{E}^i \in \mathbb{R}^{p_1 \times p_2}$ is element-wise independent idiosyncratic noise. The data generation procedure is as follows. First, the mean matrix $M_F \in \mathbb{R}^{r_1 \times r_2}$ has entries drawn independently from $\operatorname{Uniform}(0, 0.1)$. The scale (standard deviation) of each element is set as $1.5$ times its mean, ensuring heteroscedasticity. That is, for each $i$, the factor matrix is generated as $$
        \mathbf{F}^i = M_F + \boldsymbol{\varepsilon}^i, \quad \text{where } \boldsymbol{\varepsilon}^i_{jk} \sim \mathcal{N}\left(0, (1.5 \cdot [M_F]_{jk})^2\right).$$
Then, the row loading matrix $R \in \mathbb{R}^{p_1 \times r_1}$ and the column loading matrix $C \in \mathbb{R}^{p_2 \times r_2}$ are generated with entries independently sampled from $\mathcal{N}(0, 1)$, leading to dense and unstructured loadings across both modes. Finally, the heterogeneous noise matrix $\mathbf{E}^i$ has independent entries with $\mathbf{E}^i_{jk} \sim \mathcal{N}(0, \sigma^2\mathbf{z}^2)$, where $\mathbf{z} \sim \mathcal{N}[0,\text{unif}(0,2)]$, and $\sigma$ is the global noise scale.

The score function is approximated using the following neural network structures for comparison: 1. the original \texttt{Conv-UNet}, which is a convolutional encoder-decoder network designed to capture local structures in images \citep{ho2020denoising,song2020score,chen2024overview}; 2. the proposed \texttt{Tucker-UNet} with random initialization for $U_d$ in the Tucker encoder and decoder (cold-start); 3. the warm-start \texttt{Tucker-UNet} uses Tucker PCA estimators, in this case, the iterated projection estimators from \cite{Yu2021Projected}, to serve as the initial linear encoder and decoder layers of the network that can be trained via score matching; 4. the fixed(-PCA) \texttt{Tucker-UNet} , whose linear encoder and decoder layers are initialized using the Tucker PCA estimators and remain fixed during training. We set $p_1=p_2=64$, $n_{\text{train}}=4096$, $r_1=r_2=8$, and chose $\sigma\in\{0.5,1\}$. We train $300$ epochs for each diffusion model and then generate $n_{\text{gen}} = 2048$ samples, i.e., $\{\tilde{\mathbf{X}}_{t_0}^i\}_{i=1}^{n_{\text{gen}}}$, to evaluate the accuracy of the learned tensor distributions. To comprehensively evaluate the performance of the competing model implementations, we adopt three complementary metrics.

\begin{table}
\centering
\caption{Quantitative comparison of tensor generation models on the synthetic datasets after 300 training epochs. For the \texttt{Tucker-Unet}s, we report the mean performance and the corresponding standard deviation under 5 different simulation and training seeds. As for the more time-consuming \texttt{Conv-Unet}, we only report the results of a typical experiment.}
\label{tab:sim_results}

\begin{subtable}{\textwidth}
\centering
\caption{Noise level $\sigma=0.5$, $\quad$ $(n_{\text{train}},c,p,q)=$\datasetshape{4096, 1, 64, 64}}
\resizebox{0.9\textwidth}{!}{%
\begin{tabular}{lccccc}
\toprule
Score structure & $\mathcal{D}(U_{\tilde{R}},U_{R})$ & $\text{CFD}$ & {Train time} & {Gen time} \\
(300 epochs)        & subspace $\downarrow$ & core $\downarrow$ & (per epoch) & ($n_{\text{gen}}=2048$) \\
\midrule
\texttt{Conv-Unet} &  0.0288  & 6.193  & 45.69 s  & 23028.21 s \\
\texttt{Tucker-Unet} (cold)  & 0.0272 $\pm$ 0.0154  & 2.915 $\pm$ 0.169 & 14.23 s & 263.89 s \\
\texttt{Tucker-Unet} (warm)  &  \bf{0.0096 $\pm$ 0.0013}  & \bf{1.857 $\pm$ 0.302}  & 14.17 s  & 264.96 s \\
\texttt{Tucker-Unet} (fixed)  & 0.0111 $\pm$ 0.0012   & 1.876 $\pm$  0.341 & 14.13 s & 265.64 s \\
\bottomrule
\end{tabular}
}
\end{subtable}

\vspace{1.5em}

\begin{subtable}{\textwidth}
\centering
\caption{Noise level $\sigma=1$, $\quad$ $(n_{\text{train}},c,p,q)=$\datasetshape{4096, 1, 64, 64}}
\resizebox{0.9\textwidth}{!}{%
\begin{tabular}{lccccc}
\toprule
Score structure & $\mathcal{D}(U_{\tilde{R}},U_{R})$ & $\text{CFD}$ & {Train time} & {Gen time} \\
(300 epochs)        & subspace $\downarrow$ & core $\downarrow$ & (per epoch) & ($n_{\text{gen}}=2048$) \\
\midrule
\texttt{Conv-Unet}    &  0.0458  & 9.192  & 45.25 s  & 23010.73 s \\
\texttt{Tucker-Unet} (cold)  & 0.0607 $\pm$  0.0395 & 3.215 $\pm$ 0.589 & 14.20 s  & 269.97 s \\
\texttt{Tucker-Unet} (warm)  &  \bf{0.0220 $\pm$  0.0027} & 2.028 $\pm$ 0.363   & 14.15 s  & 265.01 s \\
\texttt{Tucker-Unet} (fixed) &  0.0268 $\pm$  0.0032  & \bf{2.024 $\pm$ 0.291}  & 14.11 s & 263.17 s \\
\bottomrule
\end{tabular}
}
\end{subtable}
\end{table}

First, the accuracy of the generated low-dimensional subspaces: we report the results only for $R$ due to symmetry, as $C$ is almost identical. Acquire $U_{R}$ by column orthogonalizing $R$. We can estimate $U_{R}$ by applying the iterated projection estimators from \cite{Yu2021Projected} to any generated tensor dataset $\{\tilde{\mathbf{X}}_{t_0}^i\}_{i=1}^{n_{\text{gen}}}$, denoted as $U_{\tilde{R}}$. We then report the scaled projection metric to reflect the subspace quality of the generated tensors:
\begin{equation*}
	\mathcal{D}(U_{\tilde{R}},U_{R})=(2r_1)^{-1/2}\left\|U_{\tilde{R}}U_{\tilde{R}}^{\top}-U_{R}U_{R}^{\top}\right\|_F,
\end{equation*}
where $U_{\tilde{R}}U_{\tilde{R}}^{\top}$ (and $U_{R}U_{R}^{\top}$) is the orthogonal projection matrix onto the column space of $U_{\tilde{R}}$ (and $U_{R}$). As such, $\mathcal{D}(U_{\tilde{R}},U_{R})$ is always between $0$, corresponding to $\text{span}(U_{\tilde{R}})= \text{span}(U_{R})$, and $1$, corresponding to orthogonal $\text{span}(U_{\tilde{R}})$ and $\text{span}(U_{R})$.

Second, the core distribution fidelity using a Frech\'{e}t distance on the projected core: we first map the tensors from the training dataset and each generated dataset to the core space using the true subspaces:
\begin{equation*}
\mathbf{f}^i= \operatorname{vec}\left(U_{R}^{\top} \mathbf{X}_0^iU_{C}  \right) \in \mathbb{R}^{r_1r_2},\quad \tilde{\mathbf{f}}^i= \operatorname{vec}\left(U_{R}^{\top} \mathbf{\tilde{X}}^i_{t_0}U_{C}  \right) \in \mathbb{R}^{r_1r_2}.
\end{equation*}
Let $\mathcal{F}_{\text{train}} = \{\mathbf{f}^i\}$ and $\mathcal{F}_{\text{fake}} = \{\tilde{\mathbf{f}}^i\}$ denote the sets of vectorized cores from the training data and the generated tensors, respectively. We compute the core Frech\'{e}t distance (CFD) as
\begin{equation}
    \mathrm{CFD}(\mathcal{F}_{\text{train}}, \mathcal{F}_{\text{fake}}) = \left\|\mu_{\text{train}} - \mu_{\text{fake}}\right\|^2 + \mathrm{Tr}\left(\Sigma_{\text{train}} + \Sigma_{\text{fake}} - 2 (\Sigma_{\text{train}} \Sigma_{\text{fake}})^{1/2}\right),
\end{equation}
where $\mu_{\cdot}$ and $\Sigma_{\cdot}$ are the empirical means and covariances of $\mathcal{F}_{\text{train}}$ and $\mathcal{F}_{\text{fake}}$, respectively, and $\|\cdot\|$ is the vector $\ell_2$ norm. A smaller CFD indicates higher core distribution fidelity.

Third, the computation time of both the training and tensor generation: we record the average time required to train each epoch and to generate a fixed number of samples (e.g., $n_{\text{gen}}=2048$) for all models \footnote{ All computations are made using a single NVIDIA RTX 4070 Ti SUPER chip for fair comparisons.}. The reported data generation time includes only the backward processes, excluding data loading and post-processing. Due to the dimensionality reduction architecture of our model, we expect significantly lower training and generation latency compared to the baseline model that does not employ Tucker dimension reduction.

In Table \ref{tab:sim_results}, we report the numerical results concerning the synthetic datasets. Notably, the Tucker dimension reduction significantly reduces both the training time (by three times) and the generation time (by approximately one hundred times) using a consumer-grade GPU, demonstrating its strong potential for deployment on widely accessible hardware. In addition, the two training schemes of \texttt{Tucker-Unet}s with prior Tucker PCA information yield similar outcomes, as they generate tensors of both higher subspace quality and higher core distribution fidelity, suggesting the data-efficient nature of \texttt{Tucker-Unet}s in learning high-dimensional tensor distributions. On the other hand, the cold-start \texttt{Tucker-Unet} generates tensors with lower subspace and core quality, which is foreseeable as no prior subspace information is injected. Finally, the warm-start \texttt{Tucker-Unet} has slightly better performance than the fixed(-PCA) \texttt{Tucker-Unet}. This might be due to the additional degrees of freedom in the Tucker encoder and decoder that can align with other trainable modules, such as the convolutional FiLM layer or the final skip connection residual block.

\section{Real Data Cases}\label{sec:real}
In this section, we demonstrate the potential practicability of the proposed Tucker diffusion model through several real-world tensor datasets, including two molecular datasets, a high-quality medical image dataset, and a New York City taxi trip dataset.

\subsection{Molecular generation}

We apply the tensor diffusion model to two real-world molecular datasets, PROTEINS \citep{borgwardt2005protein,morris2020tudataset} and PRC\_FM \citep{helma2001predictive,kriege2012subgraph}. Following \cite{wen2024tensor,chen2025distributed}, topological data analysis (TDA) is employed to encode each graph into an order-3 tensor of dimensions $2\times 50\times 50$ for PROTEINS and $5\times 50\times 50$ for PRC\_FM, which is composed of $50\times 50$ persistence images constructed by several filtration functions. We split the datasets into a training set ($80\%$) and a test set ($20\%$). Due to the small sample size of the PRC\_FM dataset (only $164$ training samples), we only consider channel-wise generation and reshape the PRC\_FM training samples from shape $(164,5,50,50)$ into $(820,1,50,50)$.

Note that for real data, we cannot observe the ``true subspaces'' as in Section \ref{sec:syn}. As such, to evaluate the subspace quality of the generated dataset, we report the test set reconstruction error. Let $(\widehat{A}_1, \dots, \widehat{A}_D)$ denote the estimated loading matrices obtained by applying HOOI \citep{zhang2018tensor,Yu2021Projected} to the generated dataset. Each $\widehat{A}_d \in \mathbb{R}^{p_d \times r_d}$ has orthonormal columns and estimates the true mode-$d$ loading space. We then project the real test set $\{\mathbf{X}_{\mathrm{test}}^i\}_{i=1}^{n_{\mathrm{test}}}$ onto the estimated Tucker subspace:
\[
    \hat{\mathbf{F}}^i_{\mathrm{test}} = \mathbf{X}_{\mathrm{test}}^i \times_{d=1}^D \widehat{A}_d^\top, \quad i \in [n_{\mathrm{test}}],
\]
and reconstruct each sample as $\hat{\mathbf{X}}_{\mathrm{test}}^i = \hat{\mathbf{F}}^i_{\mathrm{test}} \times_{d=1}^D \widehat{A}_d$. The normalized test reconstruction error is defined as:
\begin{equation}
    \text{RE}_{\text{test}} = 
    \frac{
        \sum_{i=1}^{n_{\mathrm{test}}} 
        \left\| \mathbf{X}_{\mathrm{test}}^i-
           \hat{\mathbf{X}}_{\mathrm{test}}^i
        \right\|_F^2
    }{
        \sum_{i=1}^{n_{\mathrm{test}}} 
        \left\| \mathbf{X}_{\mathrm{test}}^i \right\|_F^2
    },
\end{equation}
where $\|\cdot\|_F$ denotes the Frobenius norm of a tensor. This metric quantifies the loss of information in the test data after being projected onto the estimated Tucker basis. A value closer to 0 indicates a better recovery of the underlying low-rank tensor structure. Regarding the core distributional fidelity, we project the real (either training or test) and generated data using the subspaces estimated from the training set via HOOI and report the core Frech\'{e}t distance, denoted as $\text{CFD}_{\text{train}}$ and $\text{CFD}_{\text{test}}$, in the same spirit as in Section \ref{sec:syn}.

\begin{table}
\centering
\caption{Quantitative comparison of tensor generation methods on two molecular datasets. We report the best performance among $3$ training seeds.}
\label{tab:molecule_generation_results}

\begin{subtable}{\textwidth}
\centering
\caption{PRC\_FM (channel-wise), $\quad$ $(n_{\text{train}},c,p,q)=$\datasetshape{820, 1, 50, 50}}
\resizebox{0.9\textwidth}{!}{%
\begin{tabular}{lcccccc}
\toprule
Score structure & $\text{RE}_{\text{test}}$ & $\text{CFD}_{\text{train}}$ & $\text{CFD}_{\text{test}}$ & {Train time} & {Gen time} \\
(300 epochs)        & subspace $\downarrow$ & core $\downarrow$ & core $\downarrow$ & (per epoch) & ($n_{\text{gen}}=1024$) \\
\midrule
\texttt{Conv-Unet}      & 0.0001  & 0.051  & \bf{0.068}  & 8.96 s  & 11511.00 s \\
\texttt{Tucker-Unet} (cold)  & 0.0018  & 0.275  & 0.437  & 1.25 s  & 95.38 s \\
\texttt{Tucker-Unet} (warm)  & \bf{0.0000} & \bf{0.027} & 0.153 & 1.26 s & 95.15 s \\
\texttt{Tucker-Unet} (fixed)  & \bf{0.0000} & 0.065 &0.140 & 1.25 s & 95.03 s \\
\bottomrule
\end{tabular}
}
\end{subtable}

\vspace{1.5em}

\begin{subtable}{\textwidth}
\centering
\caption{PROTEINS, $\quad$ $(n_{\text{train}},c,p,q)=$\datasetshape{532, 2, 50, 50}}
\resizebox{0.9\textwidth}{!}{%
\begin{tabular}{lcccccc}
\toprule
Score structure & $\text{RE}_{\text{test}}$ & $\text{CFD}_{\text{train}}$ & $\text{CFD}_{\text{test}}$ & {Train time} & {Gen time} \\
(300 epochs)        & subspace $\downarrow$ & core $\downarrow$ & core $\downarrow$ & (per epoch) & ($n_{\text{gen}}=1024$) \\
\midrule
\texttt{Conv-Unet}      & 0.0009  & 1.256  & 1.269  & 5.728 s  & 11639.84 s \\
\texttt{Tucker-Unet} (cold)  & 0.0072  & 1.070  & 1.248  & 0.86 s  & 103.78 s \\
\texttt{Tucker-Unet} (warm)  & \bf{0.0000}  & 1.129  & \bf{1.198}  & 0.86 s  & 104.04 s \\
\texttt{Tucker-Unet} (fixed)  & 0.0002 & \bf{1.051} &1.301 & 0.85  s & 103.88 s \\
\bottomrule
\end{tabular}
}
\end{subtable}

\end{table}

The molecular generation results are reported in Table \ref{tab:molecule_generation_results}. Similar to the results in Table \ref{tab:sim_results}, we observe a significant computational advantage of the low-dimensional \texttt{Tucker-Unet} over the high-dimensional \texttt{Conv-Unet}. In addition, in terms of generated subspace quality, the cold-start \texttt{Tucker-Unet} has relatively unsatisfactory performance compared to the other versions that encode prior Tucker PCA information. In practice, we suggest using the warm-start \texttt{Tucker-Unet}, which can accelerate the training process and also ensure numerical stability.

\subsection{OrganAMNIST}

In this section, we consider learning high-quality medical image distributions from MedMNIST \citep{yang2023medmnist}. We train the Tucker diffusion model using OrganAMNIST ($128\times 128$), which comes from the liver tumor segmentation benchmark (LiTS) \citep{bilic2023liver}.

\begin{figure}
  \centering
  \begin{minipage}[t]{0.38\linewidth}
    \centering
    \includegraphics[width=\linewidth]{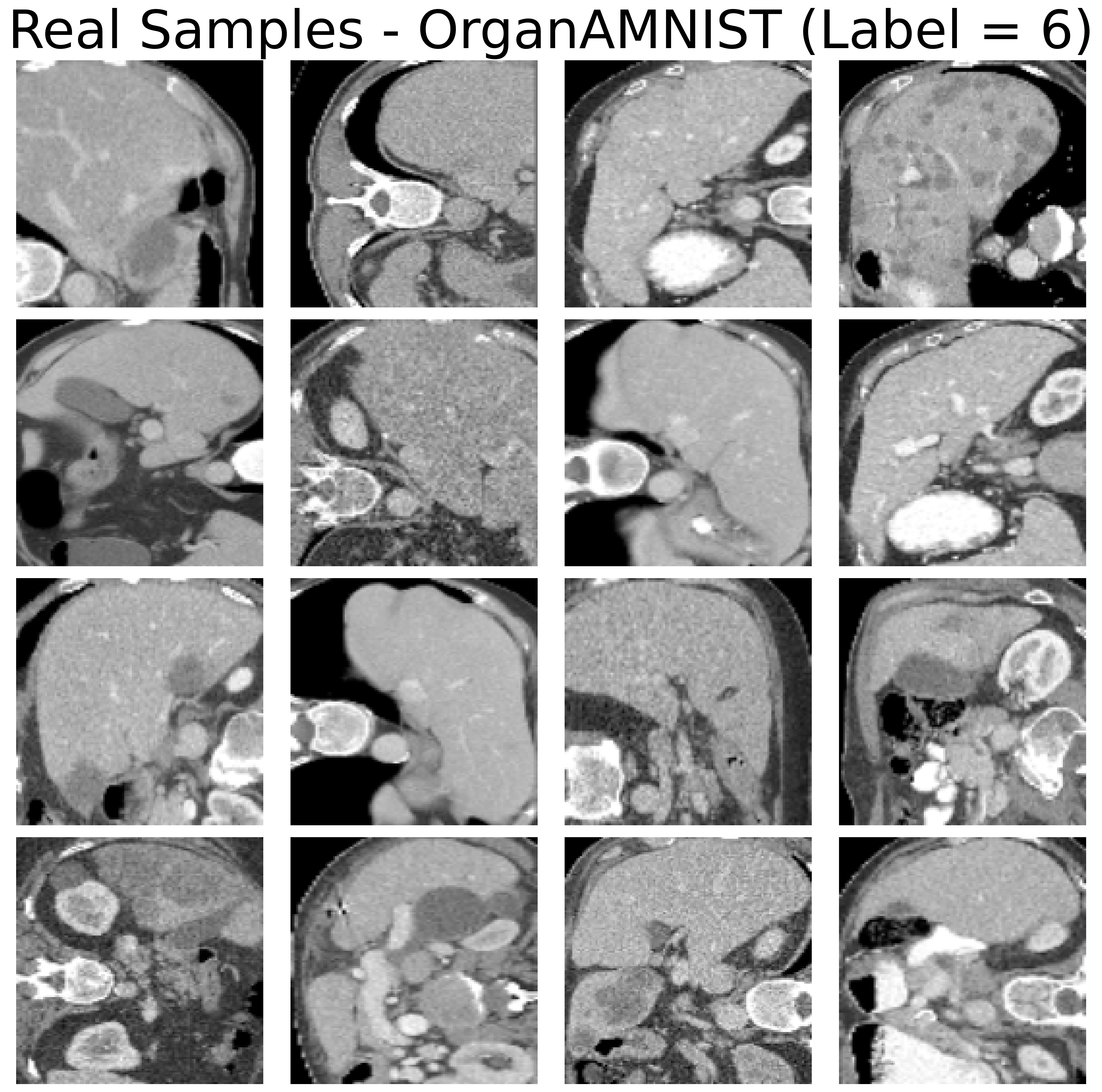}
  \end{minipage}
  \begin{minipage}[t]{0.414\linewidth}
    \centering
    \includegraphics[width=\linewidth]{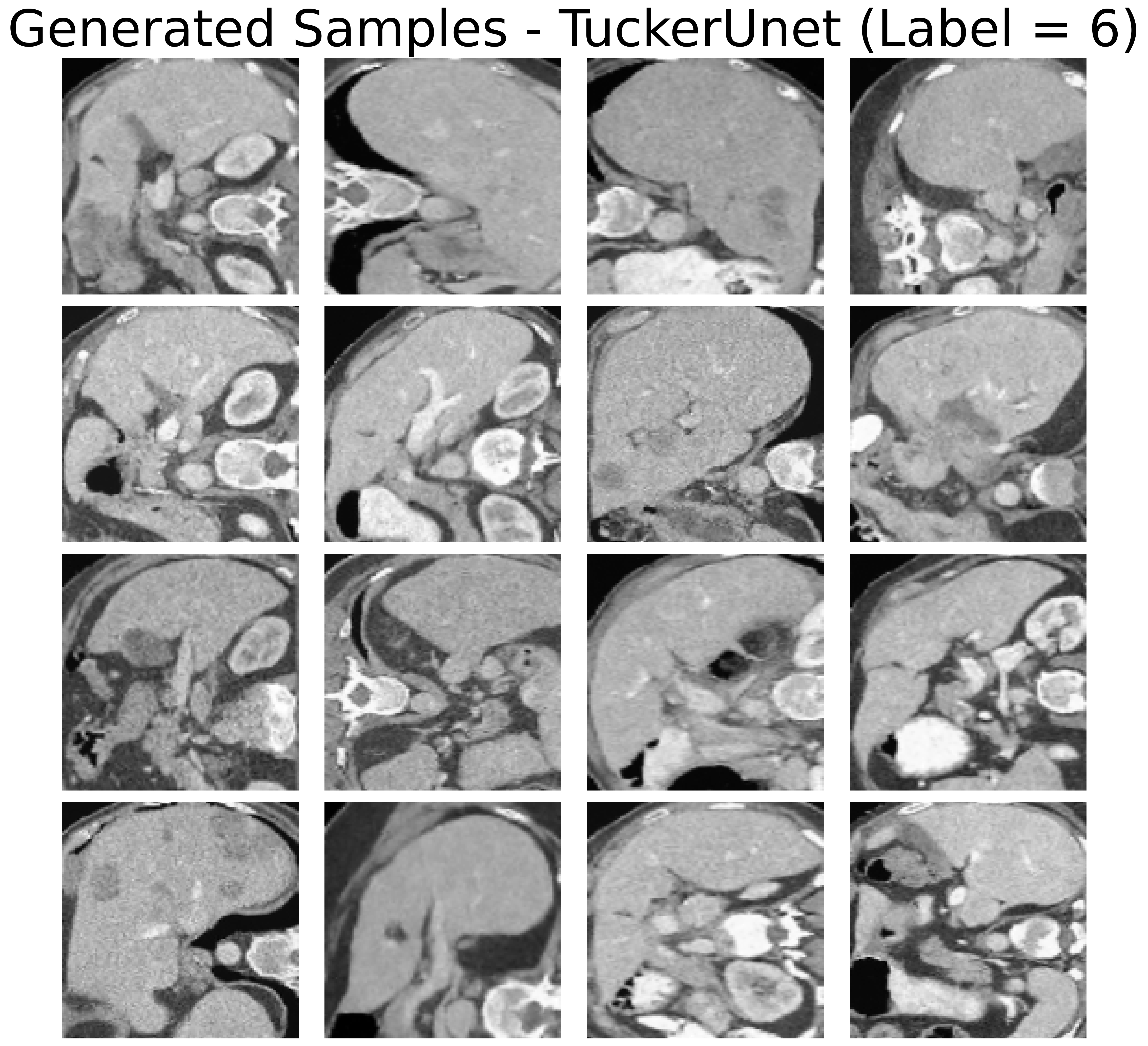}
  \end{minipage}
  \begin{minipage}[t]{0.38\linewidth}
    \centering
    \includegraphics[width=\linewidth]{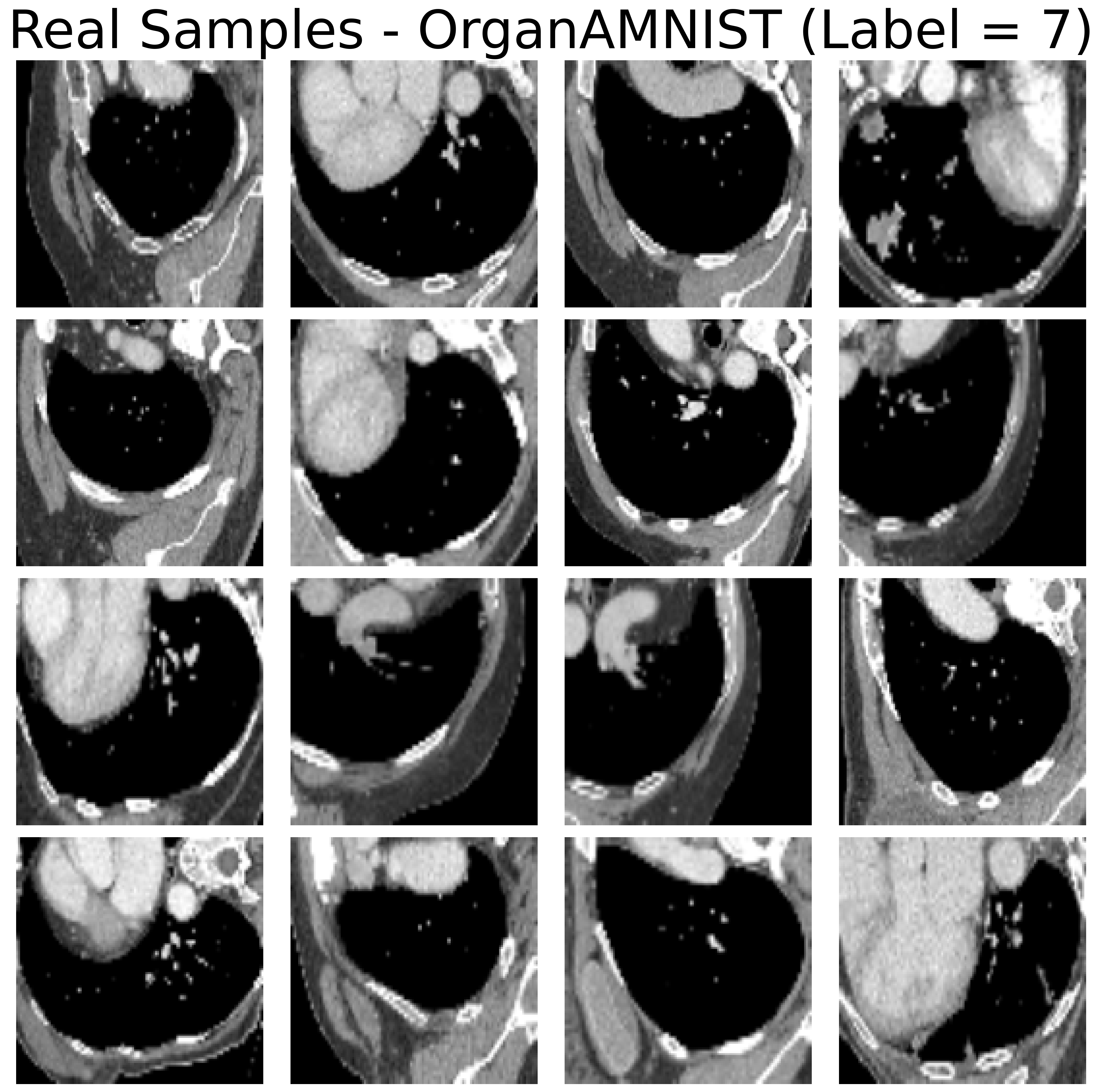}
  \end{minipage}
  \begin{minipage}[t]{0.414\linewidth}
    \centering
    \includegraphics[width=\linewidth]{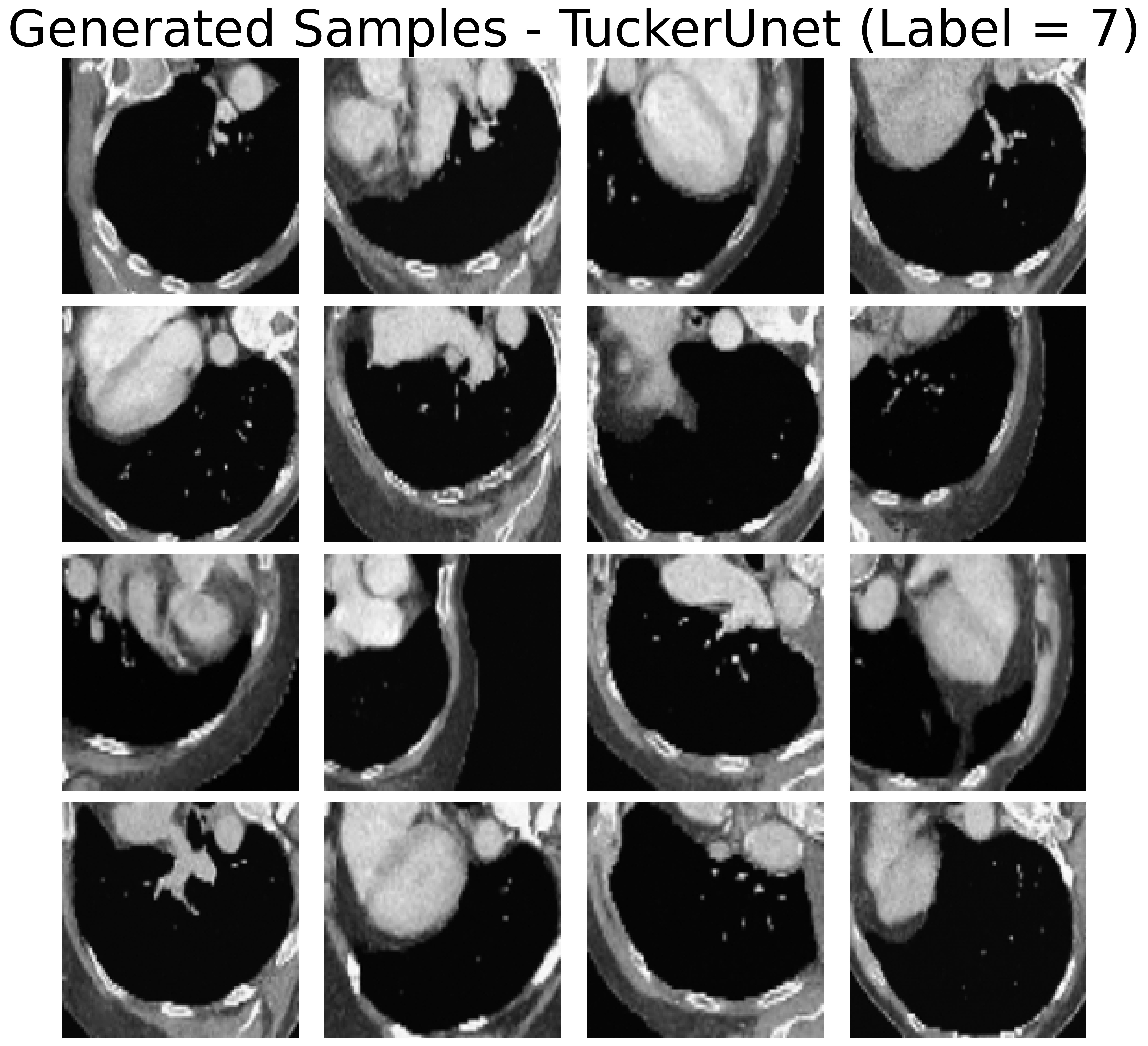}
  \end{minipage}
  \caption{Real samples from the OrganAMNIST dataset and the generated samples using the low-Tucker-rank diffusion. The Tucker diffusion model is able to retrieve the high-dimensional distribution of these medical images with significantly smaller training and sampling costs.}
  \label{fig:med}
\end{figure}

In Figure \ref{fig:med}, we present the real samples from the OrganAMNIST dataset (labels = 6 and 7), along with the generated samples from trained Tucker diffusion models using the warm-start \texttt{Tucker-Unet}. The diffusion models take $r_d=32$ and are trained for $500$ epochs. For label = 6, the training set is of shape $(n_{\text{train}},c,p,q)=(5000, 1, 128, 128)$, while for label = 7, the training set is of shape $(n_{\text{train}},c,p,q)=(3919, 1, 128, 128)$. 

We can see that the Tucker diffusion model is able to retrieve the high-dimensional distribution of these medical images with relatively small sample sizes, thanks to their inherent low-dimensionality. It is also worth mentioning that it took about 15 seconds to generate the 32 images of shape $128\times 128$ in Figure \ref{fig:med} using a single NVIDIA RTX 4070 Ti SUPER. Using the same chip, it requires more than $380$ seconds to generate 32 images of shape $64\times 64$, a much easier task, using the classical \texttt{Conv-Unet} without the Tucker structure.

\subsection{New York taxi transition}

Finally, we apply the proposed method to the New York City Yellow Taxi Trip dataset, which is a trip-level repository that includes detailed information such as pickup (PU) and drop-off (DO) timestamps and locations, travel distances, itemized fare components, rate types, payment methods, and passenger counts \citep{liu2012understanding}. 

We construct daily transition pairs from January 2019 to December 2024 between $263$ city regions. For every hour $h$ in each day, we count the number of taxi trips from the $i$-th region to the $j$-th region, collected into a $(24,263,263)$ tensor. Then, we remove those regions with fewer than $1000$ trips and focus only on the top $128$ regions with higher average daily trip numbers (the sum of PU and DO numbers). The 135 regions that are left-out typically have fewer than 50 trips per day. We end up with a tensor dataset of shape $(1790,24,128,128)$, and randomly split it into a training set ($80\%$) and a test set ($20\%$).

\begin{figure}
  \centering
  \begin{minipage}[t]{0.49\linewidth}
    \centering
    \includegraphics[width=\linewidth]{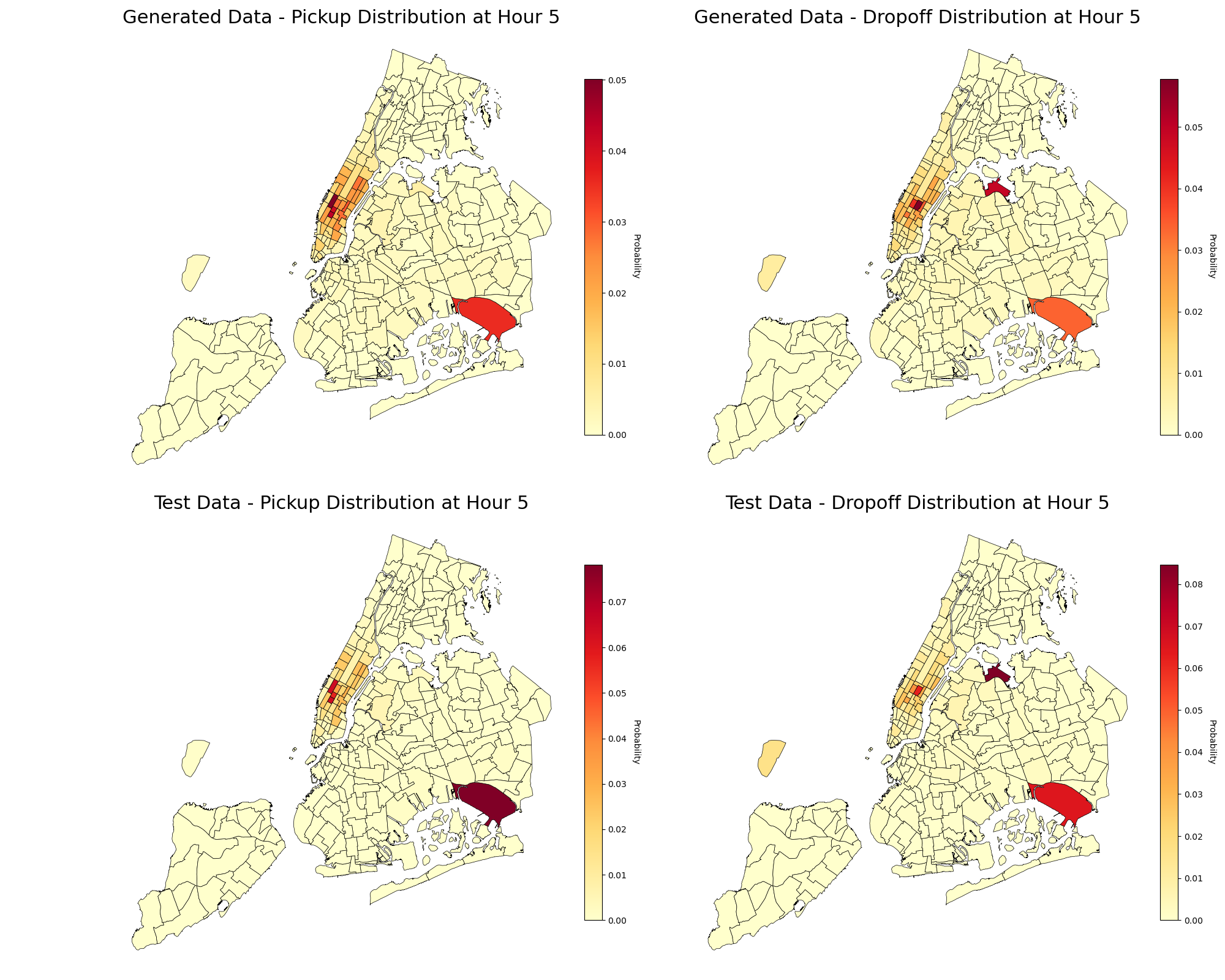}
  \end{minipage}
    \begin{minipage}[t]{0.49\linewidth}
    \centering
    \includegraphics[width=\linewidth]{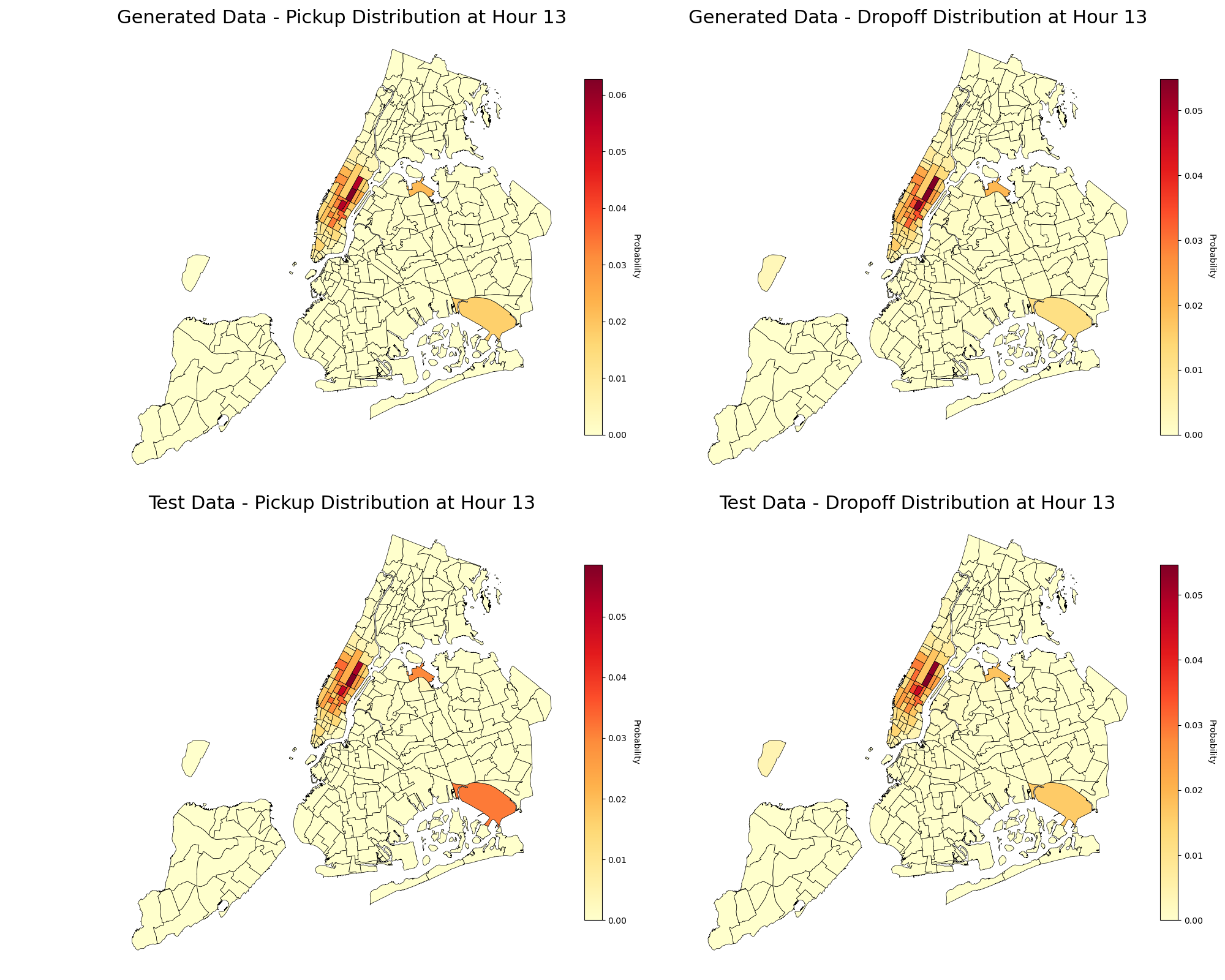}
  \end{minipage}
  \caption{Pickup (PU) and drop-off (DO) distribution patterns acquired from the generated dataset (1280 days) and the real test dataset (356 days) at different hours. The low-dimensional tensor diffusion model is able to correctly identify the airport zones and Manhattan’s major tourist, business, and hospitality districts.}
  \label{fig:taxi}
\end{figure}

As shown by \cite{zhang2019spectral,zhu2022learning}, this particular state-transition system is inherently low-rank, as the raw states can often be mapped to meta-states through factorization models on the state modes. Following their results, we set $(r_1,r_2,r_3)=(24,32,32)$ and train a warm-start \texttt{Tucker-Unet} for $500$ epochs. After that, we use the trained Tucker diffusion model to generate $n_{\text{gen}}=1280$ daily transition tensors.

In Figure \ref{fig:taxi}, for visual comparison, we compute the average hourly PU and DO marginal distributions using the generated dataset (1280 days) and the test dataset (356 days), respectively. We can see that the low-dimensional tensor diffusion model is able to correctly identify the airport zones and Manhattan’s major tourist, business, and hospitality districts. Indeed, for each hour, we can list the busiest PU and DO regions: the top 30 PU regions calculated using the generated and test tensors have an average overlap of 28.3/30 across 24 hours, while the average overlap for DO regions is 28.2/30. This indicates that the Tucker diffusion model is able to learn the spatial transition dynamics encoded in PU–DO pairs and can be further utilized to uncover latent urban mobility patterns, thereby supporting data-driven decision-making in transportation planning.
\section{Discussion}\label{sec:discussion}

In this work, we introduce the Tucker diffusion model that enables the learning and generation of tensor-valued distributions by leveraging a novel structural decomposition of the score function under the low Tucker rank assumption. Theoretically, by exploiting tensor multi-linearity, the proposed \texttt{Tucker-Unet} effectively addresses the dimensionality challenges. Empirically, experiments across synthetic and real-world datasets demonstrate the potential practicability of our method. Collectively, these advancements offer a scalable, efficient, and theoretically grounded solution for high-dimensional tensor generation.

For future directions, we can extend our work to multiple downstream tasks, such as tensor completion, addressing challenges in applications like image inpainting and recommendation systems \citep{candes2012exact,chen2020noisy,radhakrishnan2022simple,radhakrishnan2025linear}. Meanwhile, it is also interesting to develop conditional tensor generation techniques to synthesize structured data, e.g., generating realistic market scenarios conditioned on economic indicators \citep{fu2024unveil,xiu2024deep}. These extensions aim to further validate the method’s versatility and real-world impact.

\bibliographystyle{apalike}
\bibliography{main}

\end{document}